\documentclass[numberedappendix,twocolumn,twocolappendix]{openjournal}

\usepackage{amssymb,amsmath}
\usepackage{lipsum}
\usepackage{orcidlink}
\usepackage{natbib}

\usepackage{xcolor}
\usepackage{hyperref}
\hypersetup{
    unicode, 
    colorlinks=true,
    linkcolor=linkcolor,
    citecolor=linkcolor,
    filecolor=linkcolor,
    urlcolor=linkcolor,
}
\definecolor{linkcolor}{rgb}{0.0,0.3,0.5}

\graphicspath{ {./} }

\begin{document}

\title{A Comparison of Galacticus and COZMIC WDM Subhalo Populations}


\author{Jack Lonergan\orcidlink{0009-0000-5753-8918}}
\email{jacklone@usc.edu}
\affiliation{Department of Physics and Astronomy, University of Southern California}

\author{Andrew Benson\orcidlink{0000-0001-5501-6008}}
\email{abenson@carnegiescience.edu}
\affiliation{The Observatories of the Carnegie Institution for Science}

\author{Xiaolong Du\orcidlink{0000-0003-0728-2533}}
\email{xdu@astro.ucla.edu}
\affiliation{UCLA Physics and Astronomy}

\begin{abstract}
    We present a comparative analysis of warm dark matter (WDM) subhalo populations generated by the semi-analytic model {\sc Galacticus} and the COZMIC suite of dark matter-only $N$-body simulations. Using a range of thermal relic WDM particle masses (3--10 keV), we examine key summary statistics---including the subhalo mass function, spatial distribution, maximum circular velocity $V_\text{max}$, and its corresponding radius $ R_\text{max} $ ---to evaluate the consistency between these two modeling frameworks. Both models predict a suppression of low-mass subhalos correlated with decreasing WDM particle mass, and that WDM subhalos tend to have lower $V_\text{max} $ and larger $ R_\text{max} $ values than their CDM counterparts at fixed mass. While {\sc Galacticus} provides more statistically precise results due to a larger sample size, the COZMIC simulations display similar qualitative trends. We discuss how differences in halo finder algorithms, simulation resolution, and modeling assumptions affect subhalo statistics. Our findings demonstrate that {\sc Galacticus} can reliably reproduce WDM subhalo distributions seen in $N$-body simulations, offering a computationally efficient tool for exploring the implications of WDM across astrophysical phenomena.
\end{abstract}









\maketitle




\section{Introduction}
\label{introduction}

Understanding the nature of dark matter remains one of the most profound unsolved problems in modern physics. The current fiducial model of the universe is the Lambda Cold Dark Matter ($ \Lambda $CDM), which accurately predicts the observed large-scale structure of the universe, for wavenumbers $ k \lesssim 10 \text{ h}/\text{Mpc} $ \citep{frenk1985cold, aghanim2020planck, peebles2020principles}, where $ (h = H_0/100 \hbox{km/s/Mpc} )$ denotes the dimensionless Hubble parameter. At smaller, non-linear scales $ (k \gtrsim 10 \text{ h}/\text{Mpc}) $, however, tensions begin to emerge between $ \Lambda $CDM predictions for the distribution of matter and observational data. A comprehensive overview of small-scale CDM challenges and their proposed solutions can be found in \cite{bullock2017small} and \cite{del2017small}. The abundance of challenges with CDM on small scales has motivated the development of alternative dark matter models \citep{spergel2000observational, tulin2018dark, dodelson1994sterile, hu2000fuzzy, hui2017ultralight, berezhiani2016dark, bhattacharya2013two, foot2015dissipative}. 

One proposed alternative model to CDM is Warm Dark Matter (WDM). Warm dark matter assumes that dark matter is a particle with relativistic thermal velocities in the early universe. These high thermal velocities allow WDM particles to free-stream out of overdense regions in the early universe, suppressing the formation of WDM halos below their free-streaming scale \citep{bode2001halo}. Warm dark matter’s ability to suppress small-scale structure and produce halo profiles with lower concentrations has motivated it as a potential solution to specific, small-scale issues with CDM, such as the too-big-to-fail (TBTF) problem and the missing satellites problem (MSP; \citealt{viel2013warm}). Several studies  have shown that the missing satellites problem can be alleviated with baryonic processes \citep{2002MNRAS.333..177B,brooks2013baryonic, jeon2025born}, and recent discoveries of Milky Way satellites \citep{kim2018missing}—though their predictions extend to satellites below $ 10^8 M_\odot $, below the resolution limit of this work—as well as satellite counts in other galaxies \citep{nierenberg2016missing} suggest that the number of satellites is actually consistent with predictions from $\Lambda$CDM. 

Although the incorporation of baryonic physics can alleviate certain small-scale tensions within the $\Lambda$CDM model, such adjustments do not necessarily establish CDM as the definitive dark matter framework. It has been shown that while baryonic processes can help reduce the number of CDM satellites in disk galaxies, the surviving satellites remain significantly brighter than observed galaxies \citep{governato2007forming}. In disk galaxies, differing implementations of baryonic processes yield substantial variation in galaxy properties such as disk-to-total mass ratios and rotation curve peaks, which can generate contradictory outcomes to observational data. \citep{marinacci2014formation}. It should also be noted that similar baryonic physics applied to WDM models would likely produce similar trends as with CDM \citep{nierenberg2016missing}.

The primary factor that influences the free-streaming scale of warm dark matter is the mass of the warm dark matter particle, $ m_\text{WDM} $ \citep{schneider2013halo}. A wide range of methods have been used to constrain the warm dark matter particle mass, including Lyman-$ \alpha $ forest observations \citep{ballesteros2021warm, garcia2025constraining}, dwarf galaxy counts \citep{dekker2022warm}, stellar stream features \citep{banik2018probing}, strong gravitational lensing \citep{gilman2020warm}, and the cosmic reionization history \citep{tan2016constraining, lopez2017warm}. Many of these approaches utilize simulations of WDM to analyze its matter distribution through summary statistics. Different modeling approaches for WDM can lead to subtle differences in the corresponding summary statistics \citep{avila2001formation, wang2007discreteness, van2016statistics}.
Additionally, the WDM particle mass is one of multiple factors which influence the decoupling mechanism of the dark matter particle. A detailed discussion on these phenomenological differences, and how they are implemented in current simulations can be found in \cite{paduroiu2022warm}. Understanding the underlying assumptions and methodologies of each simulation is crucial for interpreting and comparing these statistical discrepancies \citep{nierenberg2016missing}.

One class of model that is able to generate halo populations more rapidly than $N$-body/hydrodynamical simulations is semi-analytic models (SAMs). SAMs simulate the formation and evolution of dark matter halos (and galaxies) by applying analytic descriptions for certain physical processes to increase the computational efficiency of the model. This computational efficiency is the primary advantage of SAMs, and enables the generation of a large number of realizations---significantly more than would be feasible with full $N$-body simulations—allowing statistical properties to be measured within a reasonable time frame \citep{henriques2009monte, benson2010galaxy, bower2010parameter}. The main drawback of using SAMs is that there is a greater degree of approximation involved compared to working with N-body simulations. Several studies have shown that SAMs are able to obtain reasonably accurate summary statistics, when compared to N-body simulations \citep{2002MNRAS.333..156B,2004MNRAS.348..811T,2005ApJ...624..505Z,2014ApJ...792...24P,2020MNRAS.498.3902Y,2023ApJ...945..159N}. SAMs have also been used to generate input data for machine learning algorithms, enabling the rapid production of statistically representative halo populations \citep{kamdar2016machine, elliott2021efficient, lonergan2025generating}.

In this paper, we compare modeling approaches of WDM dark matter subhalo populations made by the semi-analytic model {\sc Galacticus} \citep{benson2012galacticus} \footnote{\href{https://github.com/galacticusorg/galacticus}{https://github.com/galacticusorg/galacticus}, we use revision \href{https://github.com/galacticusorg/galacticus/commit/79e9402720115057c0ebaa7b765a71432bd88e62}{79e9402}.} and the COsmological ZooM-in simulations with Initial Conditions beyond CDM (COZMIC) N-body simulation suite \citep{nadler2025cozmic, an2025cozmic}. Work has already been done to compare {\sc Galacticus} CDM halo populations against simulation data \citep{yang2020new,nadler2023symphony}, as well as to compare predictions for WDM halo populations from another semi-analytic model, {\sc SASHIMI-W} \citep{ando2023sashimi}, with simulations \citep{ono2025comparison}. The results from \cite{ono2025comparison} primarily focus on comparing subhalo mass functions (SMFs) between models. In this work, we will expand on these results by comparing additional summary statistics as well as including simulations with a wider range of WDM particle masses in our analysis.

This paper is organized as follows: In Section \ref{sec:models}, we introduce the simulations and semi-analytic model {\sc Galacticus} used to generate both WDM and CDM halo populations. In Section \ref{sec:results}, we present our results in the form of summary statistics characterizing each halo population. In Section \ref{sec:dicussion}, we discuss the implications of our results, including possible sources of discrepancy and limiting features of each model. In Section \ref{sec:conclusions}, we summarize with general conclusions. 

\section{Models}
\label{sec:models}

Here, we outline the modeling frameworks used to generate $ z = 0 $ subhalo populations. While the main focus of this paper is to compare WDM subhalo populations, we also introduce CDM zoom-in N-body simulations, which will serve as a comparative reference.  Each model in this work adopts cosmological parameters: $ h = 0.7, \Omega_m = 0.286, \Omega_b = 0.047, \Omega_\Lambda = 0.714, \sigma_8 = 0.82, $ and $ n_s = 0.96 $ \citep{hinshaw2013nine}. 

\subsection{COZMIC Simulations}

We use data from the COZMIC simulation suite \citep{nadler2025cozmic}, which consists of dark matter only N-body zoom-in simulations of Milky Way (MW) mass host halos for beyond-CDM models. We specifically utilize the results for thermal-relic WDM models, where the linear matter power spectrum is characterized by the warm dark matter particle mass. (Sub-)halo populations were generated for $ m_\text{WDM} = 3,4,5,6,6.5,10 $ keV models with three halo realizations per model designed to resimulate two halos from the Milky Way-est simulation suite \citep{buch2024milky} plus one halo from Symphony. Hereafter, these models will be referenced as WDM$X$, where $ X \in \{3, 4, 5, 6, 6.5, 10 \} $. Initial conditions were generated using {\sc MUSIC} \citep{hahn2011multi}, and the simulations were run using {\sc GADGET-2} \citep{springel2005cosmological} from $ z = 99 $ down to $ z = 0 $. In the highest resolution region of each simulation, the particle mass is $ m = 4.0 \times 10^5 \mathrm{M}_\odot $, equivalent to $8192^3$ particles in the original simulation volume from which the resimulations were drawn. Requiring at least 300 particles per halo to ensure a well-converged subhalo mass function, this corresponds to a halo mass resolution of $ m_\text{res} = 1.2 \times 10^8 \mathrm{M}_\odot $. To ensure other summary statistics, such as the $ V_\text{max} $ and $ R_\text{max} $ distributions, are well-converged, a higher 2,000 particle per halo limit is imposed \citep{nadler2025cozmic}. At this limit, the minimum halo mass becomes $ m_\text{res} = 8.0 \times 10^8 \mathrm{M}_\odot $ at the fiducial resolution. To accurately resolve summary statistics beneath this mass threshold, a higher resolution resimulation of COZMIC realization Halo004 is used with a particle mass of $ m = 5.0 \times 10^4 \mathrm{M}_\odot $, enabling convergence down to $ 1.0 \times 10^8 \mathrm{M}_\odot $. WDM halos in the COZMIC simulations are identified using the {\sc RockStar} phase-space halo finder \citep{behroozi2012rockstar}. Full details of the simulation suite are described in \cite{nadler2025cozmic}.

\subsection{Symphony Simulations}

The primary aim of this work is to analyze warm dark matter subhalo populations. To assess how well {\sc Galacticus} performs in matching the COZMIC WDM simulations, we also compare {\sc Galacticus} CDM subhalo populations to CDM halos from the Symphony simulation suite---this will allow us to determine if {\sc Galacticus} performs equally well for WDM and CDM. Symphony is a set of cosmological zoom-in simulations of CDM halos in the $ 10^{11} $--$10^{15} \mathrm{M}_\odot $ range. We specifically make use of the MW mass host halos, where in the most refined region, the mass and spatial resolution are equivalent to those of a uniform simulation with $ 8192^3 $ particles in the simulation box and a simulation particle mass of $ m = 4.0 \times 10^5 \mathrm{M}_\odot $. As with the COZMIC data, a higher resolution resimulation was run for realization Halo004 with particle mass of $ 5.0 \times 10^4 \mathrm{M}_\odot $ to ensure the distributions of halo density profiles are well converged down to the $ 1.0 \times 10^8 \mathrm{M}_\odot $ halo mass scale. Among these MW mass halos, one realization, Halo023, was also resimulated by COZMIC for each WDM mass. A complete description of the simulation setup can be found in \cite{nadler2023symphony}.

\subsection{Milky Way-est Simulations}

The majority of N-body CDM halos used in this work are taken from the Symphony simulation suite. There are two additional halos used in our analysis taken from the Milky Way-est simulations \citep{buch2024milky}. Milky Way-est is a simulation suite composed of 20 cosmological cold-dark-matter-only zoom-in simulations of halos specifically tailored to match conditions of the Milky Way. In addition to a comparable host halo mass, Milky Way-est halos are also constructed to match the Milky Way's concentration and merger history. This includes an early time merger with a Gaia-Sausage-Enceladus (GSE) analog, and a subhalo with the infall and orbital properties similar to the Large Magellanic Cloud (LMC). 

Milky Way-est simulations have a particle mass in the highest resolution regions of  $ m = 4.0 \times 10^5 \mathrm{M}_\odot $, equivalent to if the original simulation box had been simulated with $ 8192^3 $ particles. Among the 20 halos in the simulation suite, Halo004 and Halo113 were designated as reference halos for constructing the corresponding WDM COZMIC analogs, and so we make use of these two Milky Way-est simulations in this work. For more details, see \cite{buch2024milky}. 

\subsection{{\sc Galacticus} Simulations}

In order to compare against the COZMIC, Symphony, and Milky Way-est simulation data, we generate subhalo populations using the {\sc Galacticus} semi-analytic framework \citep{benson2012galacticus}. In {\sc Galacticus}, subhalo populations are generated by constructing a merger tree that identifies all progenitor halos, and their sequence of merging, of a halo at $z=0$. The merger tree is constructed through a Monte Carlo (MC) process using branching rates as predicted from the Extended Press-Schechter (EPS) formalism \citep{bower1991evolution, lacey1994merger}, but modified to match the results of CDM $N$-body simulations following the algorithm of \cite{parkinson2008generating}. Once the merger tree is constructed, each halo in the tree is numerically evolved forwards in time, yielding a $ z = 0 $ subhalo population. During the evolution process, orbital properties of each halo, such as their 3D position, velocity, bound mass, and density profile are found by solving a set of differential equations that describe the physics that affects subhalos \citep{nadler2023symphony,lonergan2025generating}. Specifically, {\sc Galacticus} accounts for non-linear phenomena such as tidal heating, tidal stripping, and dynamical friction of subhalos as has previously been described and validated for CDM \citep{pullen2014nonlinear, yang2020new, benson2022tidal}. It should be noted that while {\sc Galacticus} uses a bottom up model for the structure formation of WDM halos, other studies point out that the structure formation can be more complex depending on the redshift and mass scale at which a halo forms \citep{paduroiu2022warm}. It would be interesting to consider structure formation histories alternative to the ``traditional" merger tree scenario in future work.

WDM subhalo populations in {\sc Galacticus} are derived by modifying a few key components of the corresponding {\sc Galacticus} CDM model. First, a WDM transfer function is used to account for the suppression of low-mass power due to free-streaming effects. We use the specific form given by \cite{bode2001halo}:
\begin{equation}
    \label{eq:transfer}
    T_\text{WDM}(k) = \left[ 1 + (\alpha k)^{2\nu} \right]^{-5/\nu}
\end{equation}
where $k$ is wavenumber, $ \alpha $ corresponds to the suppression scale and $ \nu = 1.049 $ from \cite{vogel2023entering}. For WDM models, we also switch to using a sharp-$k$ window function to compute $\sigma(M)$ (the root-variance in the density field) following \cite{2013MNRAS.428.1774B} to avoid the creation of spurious halos below the cut-off scale, and with a normalization parameter $a=2.5$ as advocated by that work. Since we retain a top-hat window function for CDM, to ensure that the halo mass function is unchanged on large scales, we scale\footnote{\cite{2013MNRAS.428.1774B} took a somewhat different approach to resolving this issue. They compute halo mass functions and merger rates by numerically solving the excursion set problem, and so introduced a factor into the excursion set barrier function to correct for the change in $\sigma(M)$ on large scales when using a sharp-$k$ window function. Since, in extended Press-Schechter theory, it is only the ratio of the barrier function to $\sigma^2(M)$ that appears in equations governing halo merger rates, our approach is equivalent.} $\sigma(M)$ by a factor of $0.83$.

The {\sc Galacticus} WDM model also incorporates modified halo concentrations. Specifically, we adopt the \cite{schneider2015structure} model, which determines WDM halo concentrations by matching them to those of CDM halos that share the same collapse time. At high masses, where there is little suppression of power in WDM relative to CDM, this results in concentrations that match those in CDM. At lower masses, the suppression of power in WDM models results in later collapse times, which are typical of more massive and less concentrated halos in CDM. As such, this model results in low mass WDM halos having lower concentrations than their CDM counterparts of the same mass.

Everything else in the WDM model, including the orbital physics and merger tree branching rates, is unchanged from the {\sc Galacticus} CDM model. 

For each WDM model from COZMIC, we generate 100 {\sc Galacticus} merger trees with matching host halo mass for each of the 3 simulated zoom-in host halos, for a total of 300 {\sc Galacticus} realizations per WDM model. A relatively large number of realizations were performed to ensure robust sampling of the subhalo population statistics. We also compare different {\sc Galacticus} CDM models against CDM N-body simulations, so we also generate 7 merger trees with matching host halo mass for each of the 47 different CDM MW mass halos, 45 from the Symphony simulation suite, plus 2 additional halos from Milky Way-est (giving 329 total CDM merger trees). The mass resolution for each {\sc Galacticus} merger tree was set to $ m = 3.0 \times 10^7 \mathrm{M}_\odot $---any subhalos whose bound mass falls below this threshold are removed from our calculation. This halo mass resolution was chosen such that the merger trees in {\sc Galacticus} are well converged at masses corresponding to the least massive, well-resolved halos in the COZMIC, Symphony, and Milky Way-est simulations. For details on the convergence of {\sc Galacticus} merger trees at different mass resolutions, see Appendix \ref{sec:AppA}.

\section{Results}
\label{sec:results}

In this section, we provide an overview of several summary statistic comparisons between {\sc Galacticus} and COZMIC WDM subhalo populations, along with corresponding comparisons to Symphony/Milky Way-est for CDM populations. In each set of results, CDM is represented with black curves, and WDM3, WDM4, WDM5, WDM6, WDM6.5, WDM10 models are represented by red, magenta, orange, green, blue, and purple curves, respectively. Solid curves denote data from $N$-body simulations, and dashed curves represent {\sc Galacticus} data. 

\begin{figure*}[hbt!]
    \centering
    \includegraphics[width = \textwidth]{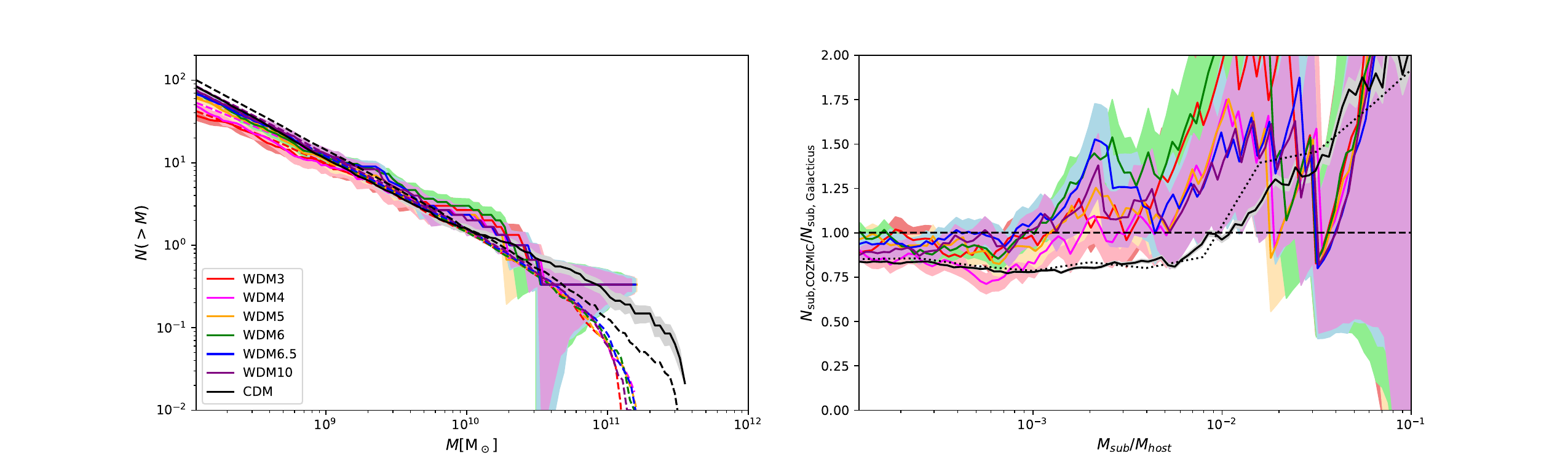}
    \caption{\emph{Left panel:} Subhalo bound mass functions for CDM (black) and WDM3, 4, 5, 6, 6.5, 10 models (red through purple). Solid curves denote the average mass function over the $N$-body simulation data from COZMIC for WDM models and Symphony/Milky Way-est for CDM. Dashed curves show the average mass function over all {\sc Galacticus} merger trees. Shaded regions around the N-body results represent uncertainty on the average mass function arising from the halo-to-halo scatter (estimated from the {\sc Galacticus} data) and the finite number of N-body realizations. \emph{Right panel:} The ratio of Symphony/Milky Way-est to {\sc Galacticus} subhalo bound mass functions (black), and ratios of COZMIC to {\sc Galacticus} subhalo bound mass functions (colored lines), as a function of the subhalo-to-host mass ratio. The dashed horizontal line shows the $ y = 1 $ line. The black dotted line indicates ratio of subhalo bound mass functions between the Symphony and Caterpillar $N$-body simulation suites.}
    \label{joint_smf}
\end{figure*}

The left panel of Figure \ref{joint_smf} shows both CDM and WDM subhalo bound mass functions at $ z = 0 $, averaged over all available realizations/merger trees, for subhalos that lie within the virial radius of the host halo and which have bound masses above the simulation mass resolution limit of $ 1.2 \times 10^8 \mathrm{M}_\odot $. For the N-body bound mass functions, we show the $ 1\sigma $ uncertainty on the average mass function arising from the halo-to-halo scatter (estimated from the {\sc Galacticus} data, for which many more realizations are available) and the finite number of N-body realizations. A clear suppression of low mass subhalos for WDM models with lower $ m_\text{WDM} $ values can be seen. 

The right panel of Figure \ref{joint_smf} shows the ratio of Symphony/Milky Way-est/COZMIC to {\sc Galacticus} subhalo bound mass functions as a function of the subhalo-to-host mass ratio, with shaded regions again showing the uncertainty on the mean due to the finite number of N-body realizations available. The horizontal dashed line represents the line of equality between $N$-body simulations and the semi-analytic model. Spatial and mass selection criteria are the same as in the left panel.

The subhalo mass function ratio curves for WDM are largely consistent with a ratio of $1$, given the uncertainty in the N-body results arising from halo-to-halo scatter and the limited number of realizations. There is a trend for the ratio to increase above the equality line for higher subhalo masses, although here the uncertainties become quite large. Importantly, there does not seem to be any systematic variation in the ratio as a function of WDM particle mass, 
which may simply be a result of the COZMIC simulations using the same set of random phases in their initial conditions, with only the amplitudes of the modes changed. The lack of $ m_\text{WDM} $ dependence in the SMF ratio curves indicates that {\sc Galacticus} is successfully reproducing the changes in the COZMIC N-body subhalo bound mass functions that arise from WDM physics.

For the CDM case (black line), the uncertainties are much smaller (due to the larger number of CDM N-body realizations), and the ratio falls significantly below 1 for low subhalo masses. A similarly low ratio was found by \cite{nadler2023symphony} when comparing the ratio of Symphony SHMFs to those from the Caterpillar simulations \citep{griffen2016caterpillar}, which is represented by the black dotted curve. The {\sc Galacticus} subhalo evolution model was calibrated against Caterpillar simulation data \citep{yang2020new}, which explains why it produces a similarly low ratio in comparison with the Symphony and COZMIC data. The CDM case also shows the trend of increasing ratio at high subhalo mass (as was previously reported by \citealt{nadler2023symphony}). Given the smaller uncertainties, this trend is significant, suggesting that {\sc Galacticus} does not accurately capture the physics of these higher mass subhalos. Furthermore, this suggests that the corresponding and very similar trend in the WDM cases is also likely a result of inaccuracies in the {\sc Galacticus} physics implementation at these higher subhalo masses, and can not simply be explained by poor statistics.

Most importantly for the present work, Figure~\ref{joint_smf} demonstrates that {\sc Galacticus} performs at least as well for WDM as it does for CDM, indicating that its implementation of WDM physics is accurate (at least for this particular summary statistic).

\begin{figure}[hbt!]
    \includegraphics[width=\linewidth]{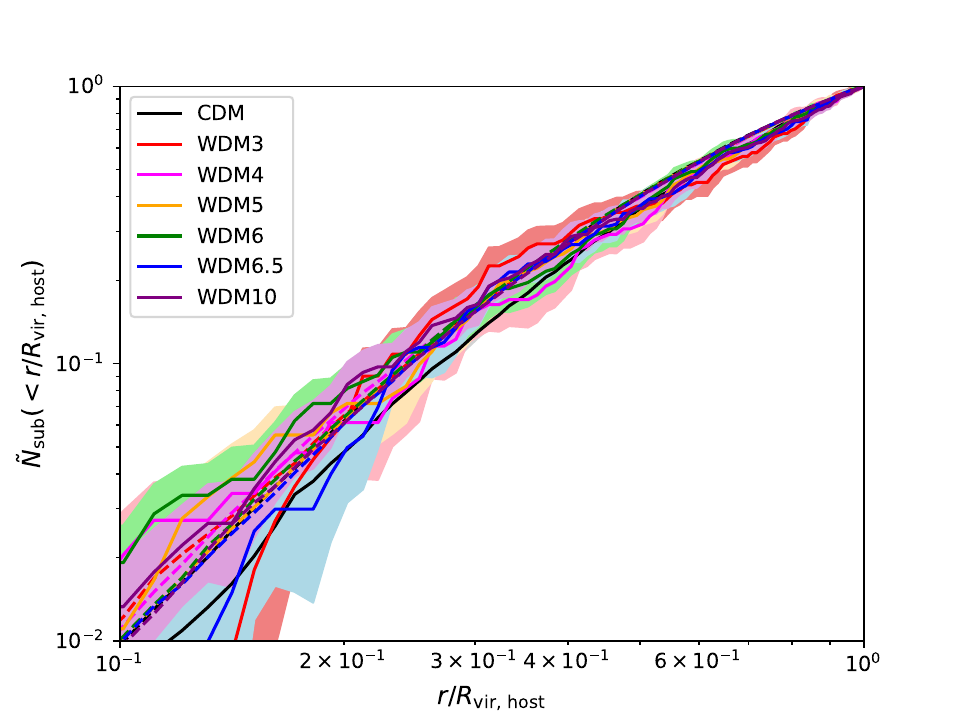}
    \centering
    \caption{Normalized radial distributions for $N$-body and {\sc Galacticus} subhalo populations. Results are shown for CDM (black) and WDM3, 4, 5, 6, 6.5, 10 models (red through purple). Solid curves denote the average radial distribution over the $N$-body simulation data from COZMIC for WDM models and Symphony/Milky Way-est for CDM. Dashed curves show the average radial distribution over all {\sc Galacticus} merger trees. Shaded regions correspond to $ 1\sigma $ uncertainties on the mean of the N-body data rising from halo-to-halo variance.}
    \label{spatial}
\end{figure}

Figure \ref{spatial} shows normalized cumulative radial distributions of subhalos for both COZMIC and {\sc Galacticus} subhalos populations, in addition to CDM results. As with the halo mass functions, this figure restricts subhalos to having masses greater than $ 1.2 \times 10^8 \mathrm{M}_\odot $. The normalized number of subhalos $ \tilde{N}_\text{sub} $ is defined by:

\begin{equation}
    \label{eq:norm_rad}
    \tilde{N}_\text{sub}(r/R_\text{vir, host}) \equiv \frac{N_\text{sub}(<r)}{N_\text{sub}(<R_\text{vir, host})}.
\end{equation}

The shaded regions show the $ 1\sigma $ uncertainties on the mean of the N-body estimates arising from halo-to-halo variance. The uncertainty on the {\sc Galacticus} results is much smaller as a result of the substantially larger number of realizations available. For WDM, the {\sc Galacticus} results remain largely enclosed within the shaded $1\sigma$ regions, suggesting agreement with COZMIC data at the level of expected statistical fluctuations. It should be noted that even at low radii, both the {\sc Galacticus} and COZMIC radial distribution curves do not scale monotonically with WDM particle mass. The results for the {\sc Galacticus} curves are statistically robust, as the difference in WDM models is larger than the statistical uncertainty around a given model. Details around this non-monotonic behavior will be discussed further in section \ref{sec:dicussion}.

\begin{figure}[hbt!]
    \includegraphics[width = \linewidth]{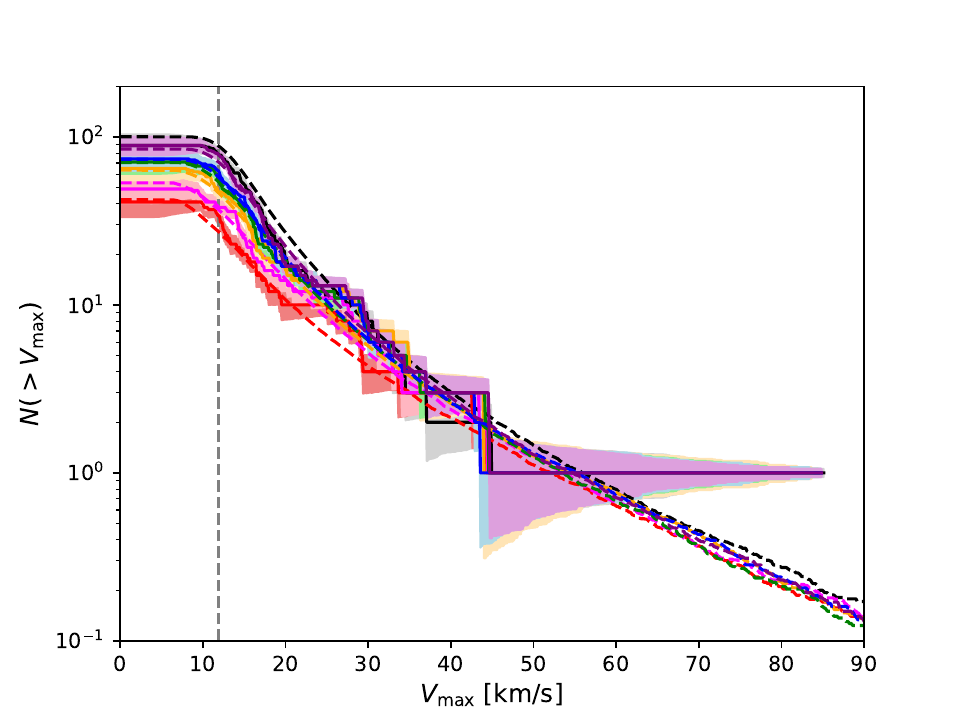}
    \centering
    \caption{Inverted cumulative distribution functions of $ V_\text{max}$ for $N$-body (solid) and {\sc Galacticus} (dashed) subhalo populations. Results are shown for both CDM (black) and WDM3 through WDM10 (red through purple) models. Shaded regions correspond to $ 1\sigma $ uncertainties on the mean of the N-body data rising from halo-to-halo variance. The vertical grey line indicates the mean $ V_\text{max} $ of halos at the high-resolution simulation threshold of $ m = 1.0 \times 10^8 \textrm{M}_\odot $ .}
    \label{vmax}
\end{figure}

We next analyze the distributions of the maximum circular velocity, $V_\text{max}$, and its corresponding radius, $R_\text{max}$, for {\sc Galacticus} and COZMIC halo populations. Figure \ref{vmax} shows inverted cumulative distributions of $ V_\text{max} $ for both {\sc Galacticus } and $ N $-body simulation data, and for both CDM and WDM models. A halo's circular velocity, $ V_\text{circ} = \sqrt{\mathrm{G}M(<r)/r} $, is proportional to the square root of the interior mass of a halo. Thus, the quantity $ V_\text{max} $ can be thought of as a proxy for the halo's mass, which does not rely on a specific definition for the edge of a halo \citep{lovell2014properties}. To ensure the distribution is well-converged, we plot the high-resolution simulation data for both WDM and CDM models, selecting halos that contain at least 2000 simulation particles, corresponding to a halo mass threshold of $ m = 1.0 \times 10^8 \mathrm{M}_\odot $. The statistic shows general trends that are present in both {\sc Galacticus} and $N$-body models. We see a general suppression at $ V_\text{max} \lesssim 40 $ km/s halos for lower mass WDM models. This is to be expected due to the small-scale suppression from WDM models, and the resulting lower halo concentration, combined with the correlation between halo mass and a halo's velocity \citep{brook2016different}. The grey dashed line in the figure denotes the average $ V_\text{max} $ of halos with masses at the $ m = 1.0 \times 10^8 \mathrm{M} _\odot $ halo mass resolution, which corresponds to a maximum circular velocity of $ V_\text{max} \approx 11.95 $ km/s. Below this limit, the $ V_\text{max} $ distributions flatten out at low velocities due to the imposed mass resolution limit. 

\begin{figure*}[hbt!]
    \includegraphics[width = \linewidth]{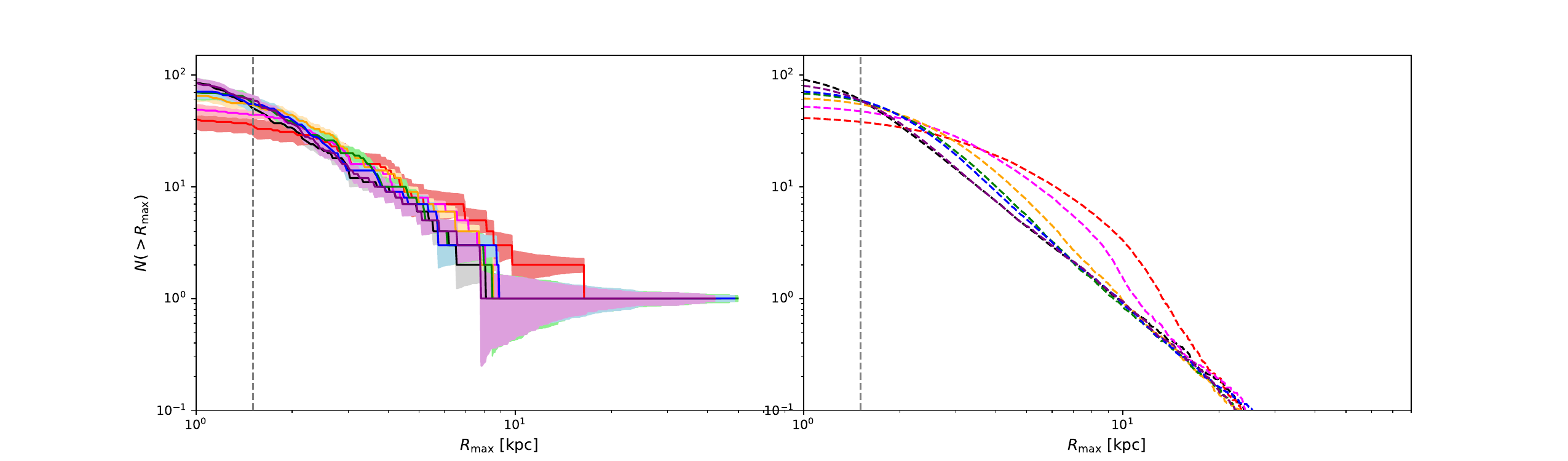}
    \centering
    \caption{Inverted cumulative distribution functions of $ R_\text{max} $ for COZMIC (left; solid curves) and {\sc Galacticus} (right; dashed curves) models. Results are shown for CDM (black) and WDM3, 4, 5, 6, 6.5, 10 models (red through purple). Shaded regions correspond to $ 1\sigma $ uncertainties on the mean of the N-body data rising from halo-to-halo variance. The grey dashed line in each panel shows the mean $ R_\text{max} $ of halos at the $ m = 1.0 \times 10^8 \mathrm{M}_\odot $ high-resolution simulation threshold.}
    \label{rmax}
\end{figure*}

We also examine the distribution of $ R_\text{max} $, the radius at which a halo's maximum circular velocity occurs. Figure \ref{rmax} depicts inverted cumulative distributions of subhalo $ R_\text{max} $ values for COZMIC models (left) and {\sc Galacticus} models (right). As in Figure \ref{vmax}, we plot the high-resolution $ N $-body simulation data to ensure the distributions in the left plot are well converged. In both {\sc Galacticus} and COZMIC lower mass WDM models exhibit a larger fraction of subhalos at higher $ R_\text{max} $. The gap is most notable in the $ R_\text{max} \sim 2$--$10 $ kpc range, although less pronounced for COZMIC subhalos compared to those from {\sc Galacticus}. Both modeling frameworks show a general trend of lower $ V_\text{max} $ and higher $ R_\text{max} $ as the WDM particle mass is decreased, highlighting that WDM halos tend to be less concentrated than their CDM counterparts \citep{lovell2014properties}. The discrepancy between different COZMIC WDM models is less significant, as only a single realization of a high-resolution simulation is available, resulting in significant statistical fluctuations. 

\begin{figure}[hbt!]
    \includegraphics[width = \linewidth]{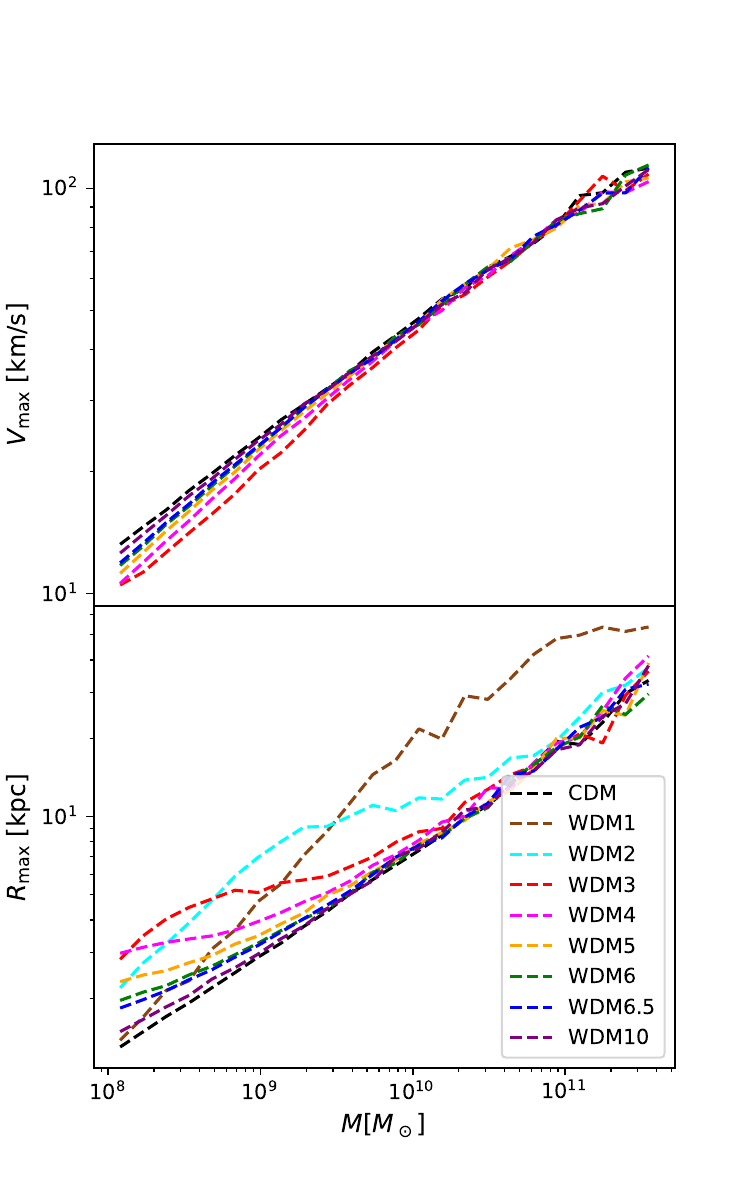}
    \centering
    \caption{Averaged subhalo $ V_\text{max} $ (top) and averaged subhalo $ R_\text{max} $ (bottom) versus halo mass for {\sc Galacticus} WDM3 through WDM10 + CDM models. The bottom panel shows two additional curves corresponding to WDM 1 keV (brown) and WDM 2 keV (cyan) models.}
    \label{vmax_rmax}
\end{figure}

To further understand the $ V_\text{max}, R_\text{max} $ properties of WDM subhalos, we show the average of each as a function of subhalo bound mass. The top panel of Figure \ref{vmax_rmax} depicts the average $ V_\text{max} $ of {\sc Galacticus} WDM subhalos as a function of subhalo bound mass. At masses $ \gtrsim 10^{10} \mathrm{M}_\odot $, each model produces similar $ V_\text{max} $ values. However, below this mass scale, there is an increasing suppression in $ V_\text{max} $  with decreasing WDM particle mass. The bottom panel similarly shows the averaged $ R_\text{max} $ for {\sc Galacticus} WDM subhalos. Once again,  below $ M < 10^{10} \mathrm{M}_\odot $ lower mass subhalos tend to have larger $ R_\text{max} $ for lower WDM particle masses. Since the curves in both figures are based on around 300 realizations, statistical noise due to sample size is negligible.

In the $R_\text{max} $ panel of Figure \ref{vmax_rmax}, for WDM4 through WDM10 + CDM models, there is a general trend such that a lower WDM particle mass results in a larger average value of $ R_\text{max} $. The WDM3 model (red curve) also follows this trend for subhalo bound masses above around $6 \times 10^8\mathrm{M}_\odot$. However, at lower masses, the WDM3 model trend changes, such that $R_\mathrm{max}$ increases less rapidly with decreasing subhalo bound mass and, at the mass resolution of $1.2 \times 10^8\mathrm{M}_\odot$, it slightly dips below the WDM4 curve. To gain some insight into the origins of this behavior, we consider two more extreme WDM models. These are included in the bottom panel of Figure~\ref{vmax_rmax} where we show $ R_\text{max} $ curves for {\sc Galacticus} WDM 1 keV (brown) and 2 keV (cyan) models. The WDM models show qualitatively similar behavior, with $R_\mathrm{max}$ initially increasing above the CDM expectation as subhalo bound mass is decreased, followed by a turnover and decline back toward the CDM expectation at lower masses. The WDM2 models dips below WDM3 around $ \sim 4\times 10^8 \mathrm{M}_\odot $, and WDM1 falls below WDM2 at approximately$ \sim 2 \times 10^9 \mathrm{M}_\odot $. 

\begin{figure}[hbt!]
    \includegraphics[width = \linewidth]{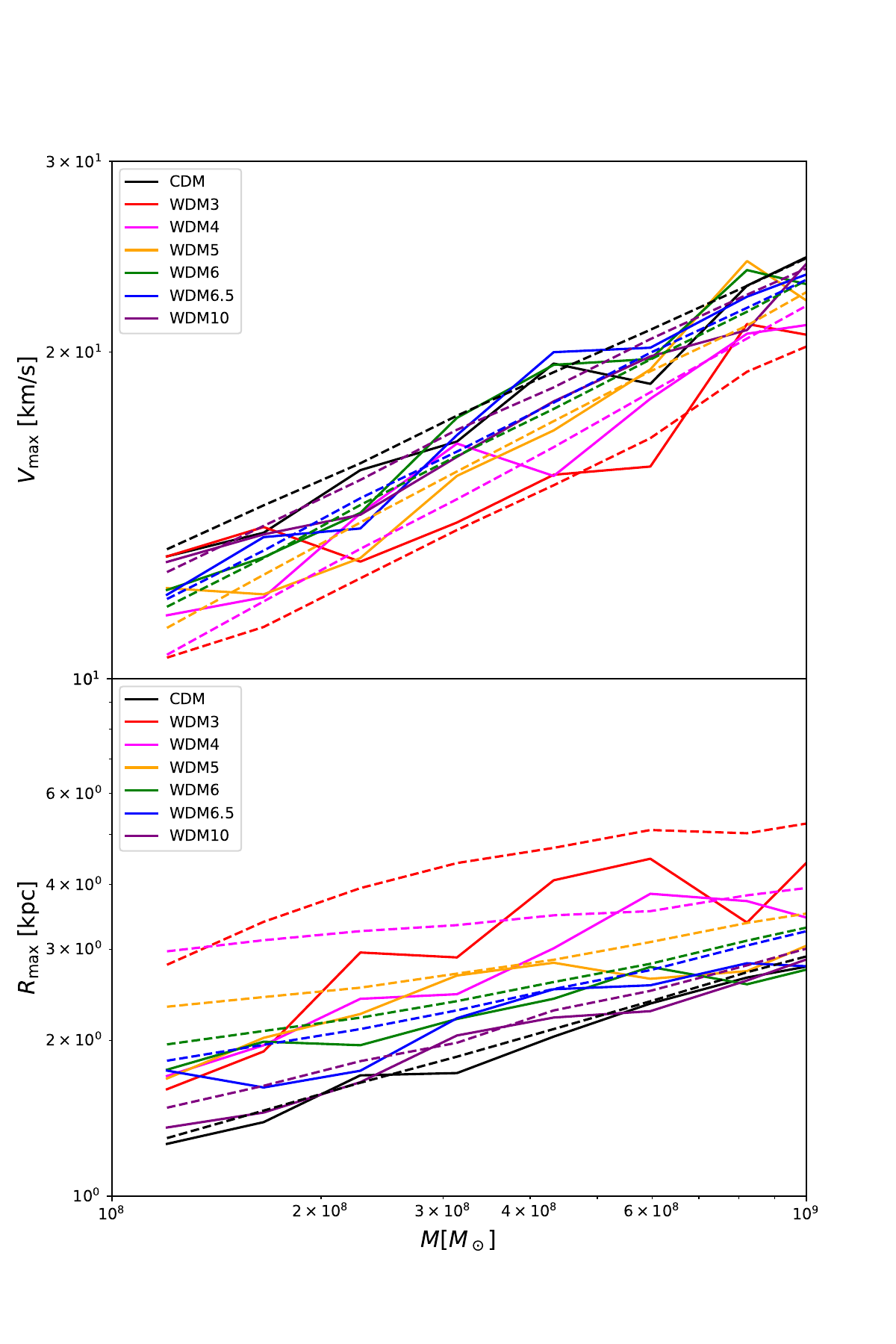}
    \centering
    \caption{Averaged subhalo $ V_\text{max} $ (top) and $ R_\text{max} $ (bottom) distributions for both {\sc Galacticus} and COZMIC subhalo populations zoomed in to the $ [10^8, 10^9] \mathrm{M}_\odot $ mass range}
    \label{zoom-in}
\end{figure}

To compare averaged $ V_\text{max}, R_\text{max} $ from {\sc Galacticus} subhalo populations against those from COZMIC, Figure \ref{zoom-in} shows the same averaged $ V_\text{max}, R_\text{max} $ values as in Figure \ref{vmax_rmax}, but now with curves for both {\sc Galacticus} and COZMIC models. The figure zooms into subhalos with masses between $ 10^8 \mathrm{M}_\odot $ to $ 10^9 \mathrm{M}_\odot $ to focus on subhalos most impacted by WDM physics. As noted by \cite{nadler2023symphony}, a 300 particle threshold in the Symphony $N$-body simulations ensures convergence of the subhalo mass function but not necessarily other summary statistics, particularly those related to the internal structure of subhalos. To robustly capture averaged $ V_\text{max} $ and $ R_\text{max} $ profiles, a stricter 2,000-particle cut is adopted. The fiducial resolution COZMIC data is plotted above its 2,000 particle mass limit of $ m = 8.0 \times 10^8 \mathrm{M}_\odot $. Below this mass scale, the higher resolution data is shown down to its 2,000 particle threshold. 

We see that although COZMIC results are noisier (due to the fact that we have data from only a single halo realization below $ m = 8.0 \times 10^8 \mathrm{M}_\odot $), there remains a general trend that lower $ m_\text{WDM} $ values lead to lower averaged $ V_\text{max} $ and higher averaged $ R_\text{max} $ values. It should be noted that the agreement between {\sc Galacticus} and COZMIC is not exact, as COZMIC tends to predict larger $ V_\text{max} $ and smaller $ R_\text{max} $ values relative to {\sc Galacticus} for a given dark matter model---particularly for the more extreme WDM models with masses at or below 5~keV. This could potentially be explained by numerical effects (as subhalos with large $ R_\text{max} $ and small $ V_\text{max} $ values are less tightly bound, which leaves them more prone to artificial tidal disruption), but may also indicate a failure of {\sc Galacticus}' physical model for these more extreme WDM particle masses. Further investigation will be needed to understand these differences.

\begin{figure}[hbt!]
    \includegraphics[width = \linewidth]{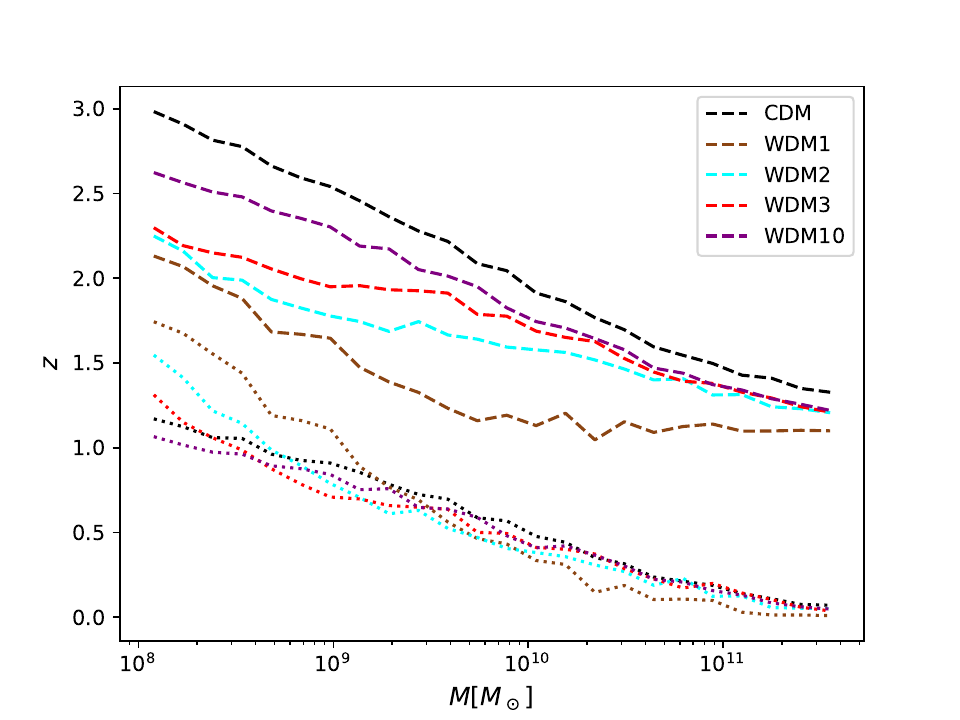}
    \centering
    \caption{Average halo redshifts as a function of halo bound mass. Dashed curves indicate averaged halo collapse redshifts, and dotted lines indicate corresponding averaged infall redshifts.}
    \label{redshift}
\end{figure}

In our {\sc Galacticus} WDM models, $R_\mathrm{max}$, at fixed subhalo mass (specifically, fixed subhalo infall mass), is determined by subhalo infall time (which sets the overall density scale of the subhalo), and collapse epoch (as defined in the \cite{schneider2015structure} model; which determines subhalo concentration). Figure \ref{redshift} shows both averaged collapse redshifts (dashed curves) and averaged infall redshifts (dotted curves), as a function of subhalo bound mass. Since a halo must form (collapse) before it can infall into another host halo, the collapse redshifts must be strictly larger than the infall redshifts. For the collapse redshift, we observe the expected trend wherein lower mass WDM models experience more significant suppression of structure formation at lower subhalo mass, leading to later collapse times. Later collapse times imply lower concentrations, and thus larger $ R_\text{max} $ values. While this trend is apparent in the WDM models, collapse redshifts in higher mass WDM models such as WDM10 do not quite converge to the CDM result even at higher subhalo masses. We attribute this to our choice to use a sharp-$k$ window function for these WDM models as opposed to the top-hat window function that we employ for CDM. Even though we scale the resulting $\sigma(M)$ in the WDM case to approximately reproduce that for CDM on large scales, this correction is not perfect and leads to the small differences in collapse epoch seen here. It should also be noted that these results represent averages over subhalos that persist to redshift $ z = 0 $. Since changes in the WDM physics influence which halos survive, these results are subject to survivor bias.

\begin{figure}[hbt!]
    \includegraphics[width=\linewidth]{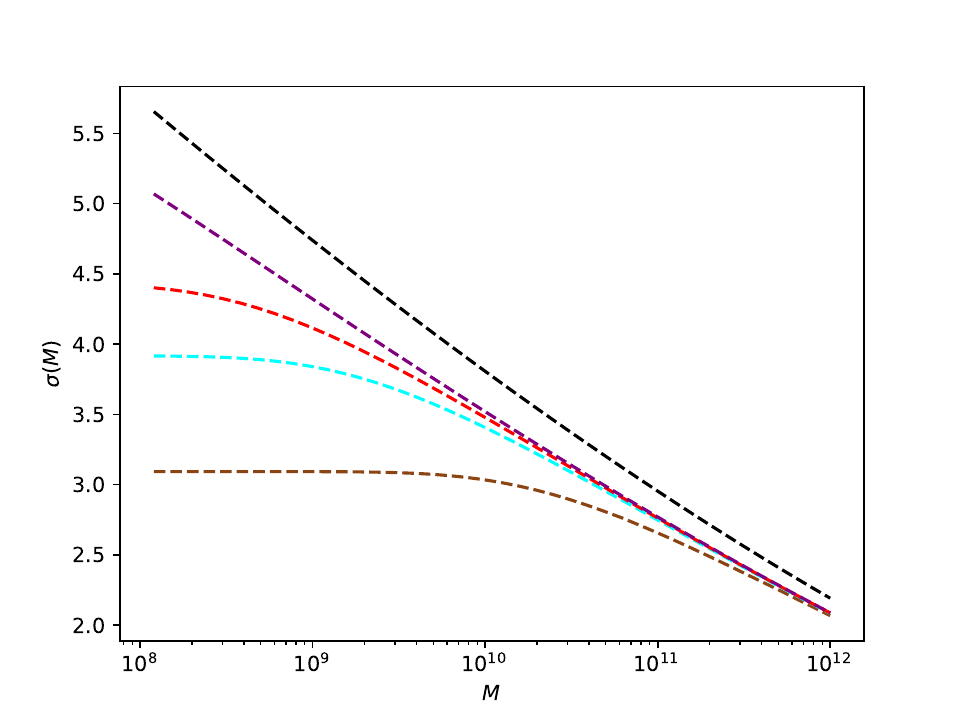}
    \centering
    \caption{Amplitude of fluctuations in the linear density field as a function of halo mass, $\sigma(M)$. Results are shown for CDM (black curve) and several WDM models, including our standard WDM10 through WDM3 models (purple through red curves), as well as WDM2 (cyan) and WDM2 (brown).}
    \label{sigma}
\end{figure}

In Figure \ref{redshift}, the WDM1 model predictably experiences the most significant suppression, resulting in lower collapse redshifts across all mass scales. However, if we examine the degree of suppression as a function of halo mass relative to the other models, some interesting behavior can be discerned. For example, the WDM1 collapse redshift curve is significantly lower than all other models in the $[10^9, 10^{10}] \mathrm{M}_\odot $ mass range; however, at the lowest masses shown, the WDM1 model is suppressed by an amount very similar to WDM2 and even WDM3 models. As noted above, the collapse redshift is constrained by the infall redshift curves (dotted lines), as halos must satisfy $ z_\text{collapse} > z_\text{infall} $. Infall redshifts correspond to when a halo first merged into a larger system and became a subhalo, and so inherently depend on the merger rates used to construct our merger trees. The merger rate in the extended Press-Schechter theory\footnote{Our actual merger tree calculations include the modifier term introduced by \cite{parkinson2008generating} to better match merger rates in N-body simulations. We do not include that here for simplicity as it does not qualitatively affect the argument that we present.} is given by \cite{1993MNRAS.262..627L}:
\begin{equation} \label{eq:merger_rate}
    \frac{\text{d}^2N}{\text{d}z_1 \text{d}M_1} = \sqrt{\frac{2}{\pi}} \frac{M_2(z)}{M_1^2} \frac{\sigma_1^2}{[\sigma_1^2 - \sigma_2^2]^{3/2}} \frac{\text{d}\delta_1}{\text{d}z_1} \left| \frac{\text{d}\log(\sigma_1)}{\text{d}\log(M_1)} \right|,
\end{equation}
where subscripts 1 and 2 refer to the progenitor and current halo (evaluated up to some redshift $ z $), respectively, and $\delta_1$ is the linear overdensity threshold for collapse of a halo. Any change in the average infall redshift of halos of fixed mass must arise from a redshift dependence in this equation\footnote{Of course, the overall normalization of the merger rate at fixed $M_1$ will be significantly different for low mass halos in WDM models because of the overall suppression of the mass function on these scales. Changing the \emph{mean} infall redshift, however, requires a change in the \emph{shape} of the merger rate as a function of redshift.}. The merger rate has a redshift dependence in the current halo mass $ M_2(z) $, the collapse threshold, $\delta_1(z)$, and the $ \sigma_2 = \sigma(M_2(z)) $ factor. Of these, $\delta_1(z)$ is strictly unchanged between CDM and all WDM models, and we have checked that $M_2(z)$ does not differ substantially across the relevant redshift range between our CDM and WDM models. Instead, the predominant change in the redshift dependence of merger rates in the WDM1 model lies in the $ \sigma_2(M_2(z)) $ factor. To see this, we show in Figure~\ref{sigma} the amplitude of fluctuations in the linear density field, $\sigma(M)$, for CDM and several WDM models.

Each model shows similar fluctuation amplitudes at higher masses; however, models begin to diverge at lower masses. The WDM1 curve is nearly constant in the mass range of MW mass subhalo populations, whereas other models are more negatively sloped in the $ [10^8, 10^{12}] \mathrm{M}_\odot $ region. Importantly, because the merger rate depends on $1/[\sigma_1^2-\sigma_2^2]^{3/2}$, as $\sigma_1$ and $\sigma_2$ become comparable, this term becomes very large, making the merger rate large.

For example, considering masses $M_1\sim 10^8\mathrm{M}_\odot$ and $M_2\sim 10^{11}$--$10^{12}\mathrm{M}_\odot$, it can be seen that, for WDM1, $\sigma_1$ and $\sigma_2$ are much closer to each other than for CDM (or for most of the other WDM models). Significantly, at higher redshifts, $M_2(z)$ decreases, making the difference between $\sigma_1$ and $\sigma_2$ smaller (and so the term in the merger rate becomes larger). Thus, for extreme WDM models, such as WDM1, merger rates are \emph{relatively} higher at higher redshifts compared to $z=0$, resulting in a bias toward higher infall redshifts. This forces a similar increase in collapse redshifts for low mass halos, which in turn leads to $R_\mathrm{max}$ trending back toward the CDM expectation for these extreme models. The same effect, albeit weaker, is present for WDM3 and causes the downturn in the average $R_\mathrm{max}$ at the lowest halo masses shown in Figure~\ref{rmax}.

\section{Discussion}\label{sec:dicussion}

We have compared several summary statistics of subhalo populations for WDM models as predicted by the COZMIC N-body simulations and the {\sc Galacticus} semi-analytic model. We find that overall, the two frameworks produce comparable subhalo populations. Within both populations, subhalo mass functions (Figure \ref{joint_smf}) show that WDM models feature a suppression of low mass halos, where the lower the $ m_\text{WDM} $ value, the more significant the suppression. The subhalo mass functions also show that, between {\sc Galacticus} and COZMIC, the amount of suppression is consistent across all different WDM models. Both {\sc Galacticus} and COZMIC models highlight that lower mass WDM halos typically have lower $ V_\text{max} $ and higher $ R_\text{max} $  than corresponding CDM models. Figure \ref{vmax_rmax} highlights the emergence of this behavior at masses $ \lesssim 10^{10} \mathrm{M}_\odot $. 

The observed differences in the summary statistics between the two frameworks can be largely attributed to the relatively small number of WDM COZMIC subhalo population realizations compared to {\sc Galacticus}. In Figure \ref{spatial}, there is a wider spread in spatial distributions among COZMIC models. However, this spread produces curves both above and below the spatial distribution of {\sc Galacticus} subhalos, and there is no apparent correlation between the spatial distribution of halos and the WDM model, indicating that the wider spatial variance in COZMIC models is due to noise associated with a small sample size (as is also apparent in Figures \ref{vmax} and \ref{rmax}). Both models show similar trends in the $ V_\text{max}, R_\text{max} $ distributions; however, the dependence on the WDM particle mass is more apparent with {\sc Galacticus} data. Future $N$-body simulations of comparable WDM halos could be integrated with COZMIC data to increase the statistical power of comparisons. It should be noted that the differences between {\sc Galacticus} and COZMIC WDM subhalo populations are comparable to the differences between {\sc Galacticus} and Symphony CDM halo populations, as the SMFs and radial distributions between the two models agree within the halo-to-halo scatter for $ M_\text{sub}/M_\text{host} > 10^{-3}$. This suggests that the source of discrepancy between {\sc Galacticus} and COZMIC stems from how {\sc Galacticus} differs from $N$-body simulations, as opposed to how {\sc Galacticus} specifically implements WDM physics. 

Certain features of the COZMIC simulations may affect the resulting summary statistics for WDM subhalo populations. $N$-body simulations rely on halo finders to extract subhalo distributions, and differences in halo finder algorithms can lead to noticeable variations in the associated summary statistics. For the results in this work, we use the {\sc RockStar} halo finder, which identifies halos by applying friends-of-friends (FoF) partitions to simulation particles based on their locations in phase space. Previous studies have shown that the choice of halo finder can produce systematic biases of up to $ \sim 50\% $ in bound mass functions, where, in particular, subhalos identified using {\sc RockStar} have significantly less bound mass than if halos were tracked using other halo finders \citep{mansfield2024symfind}. One possible avenue for future work includes comparing {\sc Galacticus} subhalo populations to those extracted from $N$-body simulations using a history-based halo finder, such as {\sc HBT-HERONS} \citep{moreno2025assessing} or SymFind \citep{mansfield2024symfind}, for more robust summary statistics.

Other factors that could influence the results for COZMIC subhalo populations include inherent limitations associated with working with cosmological zoom-in simulations. A prominent limiting feature of such simulations is the mass resolution, which imposes a minimum mass at which dark matter subhalos can be resolved. In this work, the halo mass resolution of halos is set at $ 1.2 \times 10^8 \mathrm{M}_\odot $, which corresponds to 300 simulation particles. This resolution was chosen as \cite{nadler2023symphony} pointed out that halo properties are reasonably well-converged at this limit. Despite this, halo masses around the simulation halo resolution become prone to numerical effects such as artificial tidal disruption \citep{van2018dark}. Such numerical effects have the strongest influence on lower mass halos, which is the regime where WDM predicts the largest deviation from CDM. It should be noted, however, that there is still ongoing discussion around the impact of artificial tidal disruption and other numerical artifacts in cosmological zoom-in simulations \citep{benson2022tidal, paun2025tidal, he2025artificial}.

The {\sc Galacticus} semi-analytic model also has certain features that could impact the results of its WDM subhalo populations. Subhalo evolution is dominated by non-linear physical processes, such as tidal stripping, tidal heating, and dynamical friction. {\sc Galacticus} makes analytic approximations to model these phenomena, thereby increasing computational efficiency \citep{pullen2014nonlinear}. Previous works have shown that when calibrated against CDM $N$-body simulations, {\sc Galacticus} is able to accurately reproduce corresponding summary statistics \citep{yang2020new, du2024tidal} despite its analytic approximations. 

\section{Conclusions}\label{sec:conclusions}

In this work, we compare the distributions of WDM subhalo populations of MW mass host halos from the semi-analytic framework {\sc Galacticus} and dark matter only $N$-body simulation suite COZMIC by examining an array of summary statistics for each subhalo population up to a halo mass resolution of $ 1.2 \times 10^8 \mathrm{M}_\odot $. We find that the two models produce realizations that agree within the uncertainties arising from halo-to-halo scatter and the relatively small number of N-body realizations available. Due to the free streaming length of warm dark matter, both models predict a suppression of lower mass subhalos, where the amount of suppression is inversely correlated with the WDM particle mass. Additionally, low-mass subhalos in both WDM frameworks showed statistically lower $V_\text{max}$ and larger $R_\text{max}$, respectively. 

As a semi-analytic model, {\sc Galacticus} is more computationally efficient than $N$-body simulations. It has been previously shown that {\sc Galacticus} is effective in reproducing summary statistics of CDM subhalo populations \citep{2020MNRAS.498.3902Y,nadler2023symphony}, and the results from this paper extend the capability of the SAM to warm dark matter models over a wide range of $ m_\text{WDM} $ masses. While this study provides a valuable first step, comparing {\sc Galacticus} subhalo populations with future WDM $N$-body simulations will enhance the statistical robustness of our conclusions. Having an accurate population of WDM subhalos generated through a SAM enables more computationally efficient analyses across multiple areas of astrophysics and cosmology, such as results in structure formation, gravitational lensing, observed satellite population statistics, stellar dynamics, and direct detection experiments.

\section{Acknowledgments}

We would like to thank Ethan Nadler for their assistance in accessing COZMIC, Symphony, and Milky Way-est data, as well as for their helpful discussions.

\bibliographystyle{aasjournal}

\bibliography{main}

@article{frenk1985cold,
  title={Cold dark matter, the structure of galactic haloes and the origin of the Hubble sequence},
  author={Frenk, Carlos S and White, Simon DM and Efstathiou, George and Davis, Marc},
  journal={Nature},
  volume={317},
  number={6038},
  pages={595--597},
  doi={10.1038/317595a0},
  year={1985},
  publisher={Nature Publishing Group UK London}
}

@book{peebles2020principles,
  title={Principles of physical cosmology},
  author={Peebles, Phillip James Edwin},
  year={2020},
  doi={10.1515/9780691206721},
  publisher={Princeton university press}
}

@article{bullock2017small,
  title={Small-scale challenges to the $\Lambda$ CDM paradigm},
  author={Bullock, James S and Boylan-Kolchin, Michael},
  journal={Annual Review of Astronomy and Astrophysics},
  volume={55},
  number={1},
  pages={343--387},
  doi={10.1146/annurev-astro-091916-055313},
  year={2017},
  publisher={Annual Reviews}
}

@article{spergel2000observational,
  title={Observational evidence for self-interacting cold dark matter},
  author={Spergel, David N and Steinhardt, Paul J},
  journal={Physical review letters},
  volume={84},
  number={17},
  pages={3760},
  doi={10.1103/PhysRevLett.84.3760},
  year={2000},
  publisher={APS}
}

@article{tulin2018dark,
  title={Dark matter self-interactions and small scale structure},
  author={Tulin, Sean and Yu, Hai-Bo},
  journal={Physics Reports},
  volume={730},
  pages={1--57},
  doi={10.1016/j.physrep.2017.11.004},
  year={2018},
  publisher={Elsevier}
}

@article{dodelson1994sterile,
  title={Sterile neutrinos as dark matter},
  author={Dodelson, Scott and Widrow, Lawrence M},
  journal={Physical Review Letters},
  volume={72},
  number={1},
  pages={17},
  doi={10.1103/PhysRevLett.72.17},
  year={1994},
  publisher={APS}
}

@article{hu2000fuzzy,
  title={Fuzzy cold dark matter: the wave properties of ultralight particles},
  author={Hu, Wayne and Barkana, Rennan and Gruzinov, Andrei},
  journal={Physical Review Letters},
  volume={85},
  number={6},
  doi={10.1103/PhysRevLett.85.1158},
  pages={1158},
  year={2000},
  publisher={APS}
}

@article{hui2017ultralight,
  title={Ultralight scalars as cosmological dark matter},
  author={Hui, Lam and Ostriker, Jeremiah P and Tremaine, Scott and Witten, Edward},
  journal={Physical Review D},
  volume={95},
  number={4},
  pages={043541},
  doi={10.1103/PhysRevD.95.043541},
  year={2017},
  publisher={APS}
}

@article{berezhiani2016dark,
  title={Dark matter superfluidity and galactic dynamics},
  author={Berezhiani, Lasha and Khoury, Justin},
  journal={Physics Letters B},
  volume={753},
  pages={639--643},
  doi={10.1016/j.physletb.2015.12.054},
  year={2016},
  publisher={Elsevier}
}

@article{bhattacharya2013two,
  title={Two-component dark matter},
  author={Bhattacharya, Subhadittya and Drozd, Aleksandra and Grzadkowski, Bohdan and Wudka, Jose},
  journal={Journal of High Energy Physics},
  volume={2013},
  number={10},
  pages={1--31},
  doi={10.1007/JHEP10(2013)158},
  year={2013},
  publisher={Springer}
}

@article{foot2015dissipative,
  title={Dissipative hidden sector dark matter},
  author={Foot, Robert and Vagnozzi, Sunny},
  journal={Physical Review D},
  volume={91},
  number={2},
  pages={023512},
  doi={10.1103/PhysRevD.91.023512},
  year={2015},
  publisher={APS}
}

@article{bode2001halo,
  title={Halo formation in warm dark matter models},
  author={Bode, Paul and Ostriker, Jeremiah P and Turok, Neil},
  journal={The Astrophysical Journal},
  volume={556},
  number={1},
  pages={93},
  doi={10.1086/321541},
  year={2001},
  publisher={IOP Publishing}
}

@article{viel2013warm,
  title={Warm dark matter as a solution to the small scale crisis: New constraints<? format?> from high redshift Lyman-$\alpha$ forest data},
  author={Viel, Matteo and Becker, George D and Bolton, James S and Haehnelt, Martin G},
  journal={Physical Review D—Particles, Fields, Gravitation, and Cosmology},
  volume={88},
  number={4},
  pages={043502},
  doi={10.1103/PhysRevD.88.043502},
  year={2013},
  publisher={APS}
}

@article{schneider2013halo,
  title={Halo mass function and the free streaming scale},
  author={Schneider, Aurel and Smith, Robert E and Reed, Darren},
  journal={Monthly Notices of the Royal Astronomical Society},
  volume={433},
  number={2},
  pages={1573--1587},
  doi={/10.1093/mnras/stt829},
  year={2013},
  publisher={Oxford University Press}
}

@article{garcia2025constraining,
  title={Constraining Mixed Dark Matter models with high redshift Lyman-alpha forest data},
  author={Garcia-Gallego, Olga and Ir{\v{s}}i{\v{c}}, Vid and Haehnelt, Martin G and Viel, Matteo and Bolton, James S},
  journal={Physical Review D},
  doi={10.1103/4k29-h99l},
  year={2025}
}

@article{dekker2022warm,
  title={Warm dark matter constraints using Milky Way satellite observations and subhalo evolution modeling},
  author={Dekker, Ariane and Ando, Shin’ichiro and Correa, Camila A and Ng, Kenny CY},
  journal={Physical Review D},
  volume={106},
  number={12},
  pages={123026},
  doi={10.1103/PhysRevD.106.123026},
  year={2022},
  publisher={APS}
}

@article{banik2018probing,
  title={Probing the nature of dark matter particles with stellar streams},
  author={Banik, Nilanjan and Bertone, Gianfranco and Bovy, Jo and Bozorgnia, Nassim},
  journal={Journal of Cosmology and Astroparticle Physics},
  volume={2018},
  number={07},
  pages={061},
  doi={10.1088/1475-7516/2018/07/061},
  year={2018},
  publisher={IOP Publishing}
}

@article{gilman2020warm,
  title={Warm dark matter chills out: constraints on the halo mass function and the free-streaming length of dark matter with eight quadruple-image strong gravitational lenses},
  author={Gilman, Daniel and Birrer, Simon and Nierenberg, Anna and Treu, Tommaso and Du, Xiaolong and Benson, Andrew},
  journal={Monthly Notices of the Royal Astronomical Society},
  volume={491},
  number={4},
  pages={6077--6101},
  doi={10.1093/mnras/stz3480},
  year={2020},
  publisher={Oxford University Press}
}

@article{ballesteros2021warm,
  title={How warm are non-thermal relics? Lyman-$\alpha$ bounds on out-of-equilibrium dark matter},
  author={Ballesteros, Guillermo and Garcia, Marcos AG and Pierre, Mathias},
  journal={Journal of Cosmology and Astroparticle Physics},
  volume={2021},
  number={03},
  pages={101},
  doi={10.1088/1475-7516/2021/03/101},
  year={2021},
  publisher={IOP Publishing}
}

@article{tan2016constraining,
  title={Constraining warm dark matter mass with cosmic reionization and gravitational waves},
  author={Tan, Wei-Wei and Wang, FY and Cheng, KS},
  journal={The Astrophysical Journal},
  volume={829},
  number={1},
  pages={29},
  doi={10.3847/0004-637X/829/1/29},
  year={2016},
  publisher={IOP Publishing}
}

@article{lopez2017warm,
  title={Warm dark matter and the ionization history of the Universe},
  author={Lopez-Honorez, Laura and Mena, Olga and Palomares-Ruiz, Sergio and Villanueva-Domingo, Pablo},
  journal={Physical Review D},
  volume={96},
  number={10},
  pages={103539},
  doi={10.1103/PhysRevD.96.103539},
  year={2017},
  publisher={APS}
}

@article{benson2012galacticus,
  title={Galacticus: A semi-analytic model of galaxy formation},
  author={Benson, Andrew J},
  journal={New Astronomy},
  volume={17},
  number={2},
  pages={175--197},
  doi={10.1016/j.newast.2011.07.004},
  year={2012},
  publisher={Elsevier}
}

@article{nadler2025cozmic,
  title={COZMIC. I. Cosmological Zoom-in Simulations with Initial Conditions Beyond Cold Dark Matter},
  author={Nadler, Ethan O and An, Rui and Gluscevic, Vera and Benson, Andrew and Du, Xiaolong},
  journal={The Astrophysical Journal},
  volume={986},
  number={2},
  pages={127},
  doi={10.3847/1538-4357/adceef},
  year={2025},
  publisher={IOP Publishing}
}

@article{an2025cozmic,
  title={COZMIC. II. Cosmological Zoom-in Simulations with Fractional non-CDM Initial Conditions},
  author={An, Rui and Nadler, Ethan O and Benson, Andrew and Gluscevic, Vera},
  journal={The Astrophysical Journal},
  volume={986},
  number={2},
  pages={128},
  doi={10.3847/1538-4357/adce83},
  year={2025},
  publisher={IOP Publishing}
}

@article{nadler2023symphony,
  title={Symphony: Cosmological zoom-in simulation suites over four decades of host halo mass},
  author={Nadler, Ethan O and Mansfield, Philip and Wang, Yunchong and Du, Xiaolong and Adhikari, Susmita and Banerjee, Arka and Benson, Andrew and Darragh-Ford, Elise and Mao, Yao-Yuan and Wagner-Carena, Sebastian and others},
  journal={The Astrophysical Journal},
  volume={945},
  number={2},
  pages={159},
  doi={10.3847/1538-4357/acb68c},
  year={2023},
  publisher={IOP Publishing}
}

@article{ono2025comparison,
  title={Comparison of simulations and semi-analytical model for WDM subhalo mass functions},
  author={Ono, Mizuki and Okamoto, Takashi and Ando, Shin'ichiro and Ishiyama, Tomoaki},
  journal={Publications of the Astronomical Society of Japan},
  doi={10.1093/pasj/psaf095},
  year={2025}
}

@article{hinshaw2013nine,
  title={Nine-year Wilkinson Microwave Anisotropy Probe (WMAP) observations: cosmological parameter results},
  author={Hinshaw, Gary and Larson, D and Komatsu, Eiichiro and Spergel, David N and Bennett, CLaa and Dunkley, Joanna and Nolta, MR and Halpern, M and Hill, RS and Odegard, N and others},
  journal={The Astrophysical Journal Supplement Series},
  volume={208},
  number={2},
  pages={19},
  doi={10.1088/0067-0049/208/2/19},
  year={2013},
  publisher={IOP Publishing}
}

@article{hahn2011multi,
  title={Multi-scale initial conditions for cosmological simulations},
  author={Hahn, Oliver and Abel, Tom},
  journal={Monthly Notices of the Royal Astronomical Society},
  volume={415},
  number={3},
  pages={2101--2121},
  doi={10.1111/j.1365-2966.2011.18820.x},
  year={2011},
  publisher={Blackwell Publishing Ltd Oxford, UK}
}

@article{springel2005cosmological,
  title={The cosmological simulation code GADGET-2},
  author={Springel, Volker},
  journal={Monthly notices of the royal astronomical society},
  volume={364},
  number={4},
  pages={1105--1134},
  doi={10.1111/j.1365-2966.2005.09655.x},
  year={2005},
  publisher={The Royal Astronomical Society}
}

@article{bower1991evolution,
  title={The evolution of groups of galaxies in the Press--Schechter formalism},
  author={Bower, Richard G},
  journal={Monthly Notices of the Royal Astronomical Society},
  volume={248},
  number={2},
  pages={332--352},
  doi={10.1093/mnras/248.2.332},
  year={1991},
  publisher={Oxford University Press Oxford, UK}
}

@article{lacey1994merger,
  title={Merger rates in hierarchical models of galaxy formation--II. Comparison with N-body simulations},
  author={Lacey, Cedric and Cole, Shanu},
  journal={Monthly Notices of the Royal Astronomical Society},
  volume={271},
  number={3},
  pages={676--692},
  doi={10.1093/mnras/271.3.676},
  year={1994},
  publisher={The Royal Astronomical Society}
}

@article{buch2024milky,
  title={Milky Way-est: Cosmological Zoom-in Simulations with Large Magellanic Cloud and Gaia--Sausage--Enceladus Analogs},
  author={Buch, Deveshi and Nadler, Ethan O and Wechsler, Risa H and Mao, Yao-Yuan},
  journal={The Astrophysical Journal},
  volume={971},
  number={1},
  pages={79},
  doi={10.3847/1538-4357/ad554c},
  year={2024},
  publisher={IOP Publishing}
}

@article{lovell2014properties,
  title={The properties of warm dark matter haloes},
  author={Lovell, Mark R and Frenk, Carlos S and Eke, Vincent R and Jenkins, Adrian and Gao, Liang and Theuns, Tom},
  journal={Monthly Notices of the Royal Astronomical Society},
  volume={439},
  number={1},
  pages={300--317},
  doi={10.1093/mnras/stt2431},
  year={2014},
  publisher={Oxford University Press}
}

@article{brook2016different,
  title={The different baryonic Tully--Fisher relations at low masses},
  author={Brook, Chris B and Santos-Santos, Isabel and Stinson, Greg},
  journal={Monthly Notices of the Royal Astronomical Society},
  volume={459},
  number={1},
  pages={638--645},
  doi={10.1093/mnras/stw650},
  year={2016},
  publisher={Oxford University Press}
}

@article{behroozi2012rockstar,
  title={The rockstar phase-space temporal halo finder and the velocity offsets of cluster cores},
  author={Behroozi, Peter S and Wechsler, Risa H and Wu, Hao-Yi},
  journal={The Astrophysical Journal},
  volume={762},
  number={2},
  pages={109},
  doi={10.1088/0004-637X/762/2/109},
  year={2012},
  publisher={IOP Publishing}
}

@article{moreno2025assessing,
  title={Assessing subhalo finders in cosmological hydrodynamical simulations},
  author={Moreno, Victor J Forouhar and Helly, John and McGibbon, Rob and Schaye, Joop and Schaller, Matthieu and Han, Jiaxin and Kugel, Roi},
  journal={arXiv preprint arXiv:2502.06932},
  doi={10.48550/arXiv.2502.06932
},
  year={2025}
}

@article{mansfield2024symfind,
  title={Symfind: Addressing the fragility of subhalo finders and revealing the durability of subhalos},
  author={Mansfield, Philip and Darragh-Ford, Elise and Wang, Yunchong and Nadler, Ethan O and Diemer, Benedikt and Wechsler, Risa H},
  journal={The Astrophysical Journal},
  volume={970},
  number={2},
  pages={178},
  doi={10.48550/arXiv.2308.10926},
  year={2024},
  publisher={IOP Publishing}
}

@article{van2018dark,
  title={Dark matter substructure in numerical simulations: a tale of discreteness noise, runaway instabilities, and artificial disruption},
  author={van den Bosch, Frank C and Ogiya, Go},
  journal={Monthly Notices of the Royal Astronomical Society},
  volume={475},
  number={3},
  pages={4066--4087},
  doi={10.1093/mnras/sty084},
  year={2018},
  publisher={Oxford University Press}
}

@article{paun2025tidal,
  title={Tidal adaptive softening and artificial fragmentation in cosmological simulations},
  author={Paun, Robert A Mostoghiu and Croton, Darren and Power, Chris and Knebe, Alexander and Ussing, Adam J and Duffy, Alan R},
  journal={Monthly Notices of the Royal Astronomical Society},
  pages={staf1229},
  doi={10.1093/mnras/staf1229},
  year={2025},
  publisher={Oxford University Press}
}

@article{he2025artificial,
  title={Why artificial disruption is not a concern for current cosmological simulations},
  author={He, Feihong and Han, Jiaxin and Li, Zhaozhou},
  journal={The Astrophysical Journal},
  volume={981},
  number={2},
  pages={108},
  doi={10.48550/arXiv.2408.04470},
  year={2025},
  publisher={IOP Publishing}
}

@article{benson2022tidal,
  title={Tidal tracks and artificial disruption of cold dark matter haloes},
  author={Benson, Andrew J and Du, Xiaolong},
  journal={Monthly Notices of the Royal Astronomical Society},
  volume={517},
  number={1},
  pages={1398--1406},
  doi={10.1093/mnras/stac2750},
  year={2022},
  publisher={Oxford University Press}
}

@article{pullen2014nonlinear,
  title={Nonlinear evolution of dark matter subhalos and applications to warm dark matter},
  author={Pullen, Anthony R and Benson, Andrew J and Moustakas, Leonidas A},
  journal={The Astrophysical Journal},
  volume={792},
  number={1},
  pages={24},
  doi={10.1088/0004-637X/792/1/24},
  year={2014},
  publisher={IOP Publishing}
}

@article{yang2020new,
  title={A new calibration method of sub-halo orbital evolution for semi-analytic models},
  author={Yang, Shengqi and Du, Xiaolong and Benson, Andrew J and Pullen, Anthony R and Peter, Annika HG},
  journal={Monthly Notices of the Royal Astronomical Society},
  volume={498},
  number={3},
  pages={3902--3913},
  doi={10.1093/mnras/staa2496
},
  year={2020},
  publisher={Oxford University Press}
}

@article{du2024tidal,
  title={Tidal evolution of cored and cuspy dark matter halos},
  author={Du, Xiaolong and Benson, Andrew and Zeng, Zhichao Carton and Treu, Tommaso and Peter, Annika HG and Mace, Charlie and Jiang, Fangzhou and Yang, Shengqi and Gannon, Charles and Gilman, Daniel and others},
  journal={Physical Review D},
  volume={110},
  number={2},
  pages={023019},
  doi={10.1103/PhysRevD.110.023019},
  year={2024},
  publisher={APS}
}

@article{aghanim2020planck,
      title={Planck 2018 results. VI. Cosmological parameters},
  author={Aghanim, N and others},
  journal={Astronomy and Astrophysics},
  volume={641},
  pages={A6},
  doi={10.1051/0004-6361/201833910},
  year={2020}
}

@article{brooks2013baryonic,
  title={A baryonic solution to the missing satellites problem},
  author={Brooks, Alyson M and Kuhlen, Michael and Zolotov, Adi and Hooper, Dan},
  journal={The Astrophysical Journal},
  volume={765},
  number={1},
  pages={22},
  doi={10.1088/0004-637X/765/1/22},
  year={2013},
  publisher={IOP Publishing}
}

@article{jeon2025born,
  title={Born to be Starless: Revisiting the Missing Satellite Problem},
  author={Jeon, Seyoung and Sukyoung, K Yi and Contini, Emanuele and Dubois, Yohan and Han, San and Kraljic, Katarina and Peirani, Sebastien and Pichon, Christophe and Rhee, Jinsu},
  journal={The Astrophysical Journal},
  volume={988},
  number={1},
  pages={136},
  doi={10.48550/arXiv.2506.09152},
  year={2025},
  publisher={IOP Publishing}
}

@article{kim2018missing,
  title={Missing satellites problem: completeness corrections to the number of satellite galaxies in the milky way are consistent with cold dark matter predictions},
  author={Kim, Stacy Y and Peter, Annika HG and Hargis, Jonathan R},
  journal={Physical review letters},
  volume={121},
  number={21},
  pages={211302},
  doi={10.1103/PhysRevLett.121.211302},
  year={2018},
  publisher={APS}
}

@article{nierenberg2016missing,
  title={The missing satellite problem in 3D},
  author={Nierenberg, AM and Treu, T and Menci, Nicola and Lu, Y and Torrey, Paul and Vogelsberger, M},
  journal={Monthly Notices of the Royal Astronomical Society},
  volume={462},
  number={4},
  pages={4473--4481},
  doi={10.1093/mnras/stw1860},
  year={2016},
  publisher={The Royal Astronomical Society}
}

@article{avila2001formation,
  title={Formation and Structure of Halos in a Warm Dark Matter Cosmology},
  author={Avila-Reese, Vladimir and Col{\'\i}n, Pedro and Valenzuela, Octavio and D’Onghia, Elena and Firmani, Claudio},
  journal={The Astrophysical Journal},
  volume={559},
  number={2},
  pages={516},
  doi={10.1086/322411},
  year={2001},
  publisher={IOP Publishing}
}

@article{wang2007discreteness,
  title={Discreteness effects in simulations of hot/warm dark matter},
  author={Wang, Jie and White, Simon DM},
  journal={Monthly Notices of the Royal Astronomical Society},
  volume={380},
  number={1},
  pages={93--103},
  doi={10.1111/j.1365-2966.2007.12053.x},
  year={2007},
  publisher={Blackwell Publishing Ltd Oxford, UK}
}

@article{van2016statistics,
  title={Statistics of dark matter substructure--II. Comparison of model with simulation results},
  author={van den Bosch, Frank C and Jiang, Fangzhou},
  journal={Monthly Notices of the Royal Astronomical Society},
  volume={458},
  number={3},
  pages={2870--2884},
  doi={10.48550/arXiv.1403.6835},
  year={2016},
  publisher={Oxford University Press}
}

@article{henriques2009monte,
  title={Monte Carlo Markov Chain parameter estimation in semi-analytic models of galaxy formation},
  author={Henriques, Bruno MB and Thomas, Peter A and Oliver, Seb and Roseboom, Isaac},
  journal={Monthly Notices of the Royal Astronomical Society},
  volume={396},
  number={1},
  pages={535--547},
  doi={10.1111/j.1365-2966.2009.14730.x},
  year={2009},
  publisher={Blackwell Publishing Ltd Oxford, UK}
}

@article{benson2010galaxy,
  title={Galaxy formation spanning cosmic history},
  author={Benson, Andrew J and Bower, Richard},
  journal={Monthly Notices of the Royal Astronomical Society},
  volume={405},
  number={3},
  pages={1573--1623},
  doi={10.1111/j.1365-2966.2010.16592.x},
  year={2010},
  publisher={Blackwell Publishing Ltd Oxford, UK}
}

@article{bower2010parameter,
  title={The parameter space of galaxy formation},
  author={Bower, Richard G and Vernon, I and Goldstein, Michael and Benson, AJ and Lacey, CG and Baugh, Carlton M and Cole, Shaun and Frenk, CS},
  journal={Monthly Notices of the Royal Astronomical Society},
  volume={407},
  number={4},
  pages={2017--2045},
  doi={10.1111/j.1365-2966.2010.16991.x},
  year={2010},
  publisher={Blackwell Publishing Ltd Oxford, UK}
}

@article{kamdar2016machine,
  title={Machine learning and cosmological simulations--I. Semi-analytical models},
  author={Kamdar, Harshil M and Turk, Matthew J and Brunner, Robert J},
  journal={Monthly Notices of the Royal Astronomical Society},
  volume={455},
  number={1},
  pages={642--658},
  doi={10.1093/mnras/stv2310},
  year={2016},
  publisher={Oxford University Press}
}

@article{elliott2021efficient,
  title={Efficient exploration and calibration of a semi-analytical model of galaxy formation with deep learning},
  author={Elliott, Edward J and Baugh, Carlton M and Lacey, Cedric G},
  journal={Monthly Notices of the Royal Astronomical Society},
  volume={506},
  number={3},
  pages={4011--4030},
  doi={10.1093/mnras/stab1837},
  year={2021},
  publisher={Oxford University Press}
}

@article{lonergan2025generating,
  title={Generating Dark Matter Subhalo Populations Using Normalizing Flows},
  author={Lonergan, Jack and Benson, Andrew and Gilman, Daniel},
  journal={The Open Journal of Astrophysics},
  doi={10.33232/001c.142569},
  year={2025}
}

@article{ando2023sashimi,
  title={SASHIMI-W: Semi-Analytical SubHalo Inference ModelIng for Warm Dark Matter},
  author={Ando, Shin'ichiro},
  journal={Astrophysics Source Code Library},
  pages={ascl--2302},
  doi={10.1088/1475-7516/2025/02/053},
  year={2023}
}

@article{parkinson2008generating,
  title={Generating dark matter halo merger trees},
  author={Parkinson, Hannah and Cole, Shaun and Helly, John},
  journal={Monthly Notices of the Royal Astronomical Society},
  volume={383},
  number={2},
  pages={557--564},
  doi={10.1111/j.1365-2966.2007.12517.x},
  year={2008},
  publisher={The Royal Astronomical Society}
}

@article{vogel2023entering,
  title={Entering the era of measuring sub-Galactic dark matter structure: Accurate transfer functions for axino, gravitino, and sterile neutrino thermal warm dark matter},
  author={Vogel, Cannon M and Abazajian, Kevork N},
  journal={Physical Review D},
  volume={108},
  number={4},
  pages={043520},
  doi={10.1103/PhysRevD.108.043520},
  year={2023},
  publisher={APS}
}

@article{schneider2015structure,
  title={Structure formation with suppressed small-scale perturbations},
  author={Schneider, Aurel},
  journal={Monthly Notices of the Royal Astronomical Society},
  volume={451},
  number={3},
  pages={3117--3130},
  doi={10.1093/mnras/stv1169},
  year={2015},
  publisher={Oxford University Press}
}

@article{griffen2016caterpillar,
  title={The Caterpillar project: a large suite of Milky Way sized halos},
  author={Griffen, Brendan F and Ji, Alexander P and Dooley, Gregory A and G{\'o}mez, Facundo A and Vogelsberger, Mark and O’Shea, Brian W and Frebel, Anna},
  journal={The Astrophysical Journal},
  volume={818},
  number={1},
  pages={10},
  doi={10.3847/0004-637X/818/1/10},
  year={2016},
  publisher={IOP Publishing}
}

@ARTICLE{2002MNRAS.333..177B,
       author = {{Benson}, A.~J. and {Frenk}, C.~S. and {Lacey}, C.~G. and {Baugh}, C.~M. and {Cole}, S.},
        title = "{The effects of photoionization on galaxy formation - II. Satellite galaxies in the Local Group}",
      journal = {Monthly Notices of the Royal Astronomical Society},
         year = 2002,
        month = jun,
       volume = {333},
       number = {1},
        pages = {177-190},
          doi = {10.1046/j.1365-8711.2002.05388.x},
archivePrefix = {arXiv},
       eprint = {astro-ph/0108218},
 primaryClass = {astro-ph},
       adsurl = {https://ui.adsabs.harvard.edu/abs/2002MNRAS.333..177B}
}

@ARTICLE{2002MNRAS.333..156B,
       author = {{Benson}, A.~J. and {Lacey}, C.~G. and {Baugh}, C.~M. and {Cole}, S. and {Frenk}, C.~S.},
        title = "{The effects of photoionization on galaxy formation - I. Model and results at z=0}",
      journal = {Monthly Notices of the Royal Astronomical Society},
         year = 2002,
        month = jun,
       volume = {333},
       number = {1},
        pages = {156-176},
          doi = {10.1046/j.1365-8711.2002.05387.x},
archivePrefix = {arXiv},
       eprint = {astro-ph/0108217},
 primaryClass = {astro-ph},
       adsurl = {https://ui.adsabs.harvard.edu/abs/2002MNRAS.333..156B}
}

@ARTICLE{2004MNRAS.348..811T,
       author = {{Taylor}, James E. and {Babul}, Arif},
        title = "{The evolution of substructure in galaxy, group and cluster haloes - I. Basic dynamics}",
      journal = {Monthly Notices of the Royal Astronomical Society},
         year = 2004,
        month = mar,
       volume = {348},
       number = {3},
        pages = {811-830},
          doi = {10.1111/j.1365-2966.2004.07395.x},
archivePrefix = {arXiv},
       eprint = {astro-ph/0301612},
 primaryClass = {astro-ph},
       adsurl = {https://ui.adsabs.harvard.edu/abs/2004MNRAS.348..811T}
}

@ARTICLE{2005ApJ...624..505Z,
       author = {{Zentner}, Andrew R. and {Berlind}, Andreas A. and {Bullock}, James S. and {Kravtsov}, Andrey V. and {Wechsler}, Risa H.},
        title = "{The Physics of Galaxy Clustering. I. A Model for Subhalo Populations}",
      journal = {The Astrophysical Journal},
         year = 2005,
        month = may,
       volume = {624},
       number = {2},
        pages = {505-525},
          doi = {10.1086/428898},
archivePrefix = {arXiv},
       eprint = {astro-ph/0411586},
 primaryClass = {astro-ph},
       adsurl = {https://ui.adsabs.harvard.edu/abs/2005ApJ...624..505Z}
}

@ARTICLE{2014ApJ...792...24P,
       author = {{Pullen}, Anthony R. and {Benson}, Andrew J. and {Moustakas}, Leonidas A.},
        title = "{Nonlinear Evolution of Dark Matter Subhalos and Applications to Warm Dark Matter}",
      journal = {The Astrophysical Journal},
         year = 2014,
        month = sep,
       volume = {792},
       number = {1},
          eid = {24},
        pages = {24},
          doi = {10.1088/0004-637X/792/1/24},
archivePrefix = {arXiv},
       eprint = {1407.8189},
 primaryClass = {astro-ph.CO},
       adsurl = {https://ui.adsabs.harvard.edu/abs/2014ApJ...792...24P}
}

@ARTICLE{2020MNRAS.498.3902Y,
       author = {{Yang}, Shengqi and {Du}, Xiaolong and {Benson}, Andrew J. and {Pullen}, Anthony R. and {Peter}, Annika H.~G.},
        title = "{A new calibration method of sub-halo orbital evolution for semi-analytic models}",
      journal = {Monthly Notices of the Royal Astronomical Society},
         year = 2020,
        month = nov,
       volume = {498},
       number = {3},
        pages = {3902-3913},
          doi = {10.1093/mnras/staa2496},
archivePrefix = {arXiv},
       eprint = {2003.10646},
 primaryClass = {astro-ph.CO},
       adsurl = {https://ui.adsabs.harvard.edu/abs/2020MNRAS.498.3902Y}
}

@ARTICLE{2023ApJ...945..159N,
       author = {{Nadler}, Ethan O. and {Mansfield}, Philip and {Wang}, Yunchong and {Du}, Xiaolong and {Adhikari}, Susmita and {Banerjee}, Arka and {Benson}, Andrew and {Darragh-Ford}, Elise and {Mao}, Yao-Yuan and {Wagner-Carena}, Sebastian and {Wechsler}, Risa H. and {Wu}, Hao-Yi},
        title = "{Symphony: Cosmological Zoom-in Simulation Suites over Four Decades of Host Halo Mass}",
      journal = {The Astrophysical Journal},
         year = 2023,
        month = mar,
       volume = {945},
       number = {2},
          eid = {159},
        pages = {159},
          doi = {10.3847/1538-4357/acb68c},
archivePrefix = {arXiv},
       eprint = {2209.02675},
 primaryClass = {astro-ph.CO},
       adsurl = {https://ui.adsabs.harvard.edu/abs/2023ApJ...945..159N}
}

@ARTICLE{2013MNRAS.428.1774B,
       author = {{Benson}, Andrew J. and {Farahi}, Arya and {Cole}, Shaun and {Moustakas}, Leonidas A. and {Jenkins}, Adrian and {Lovell}, Mark and {Kennedy}, Rachel and {Helly}, John and {Frenk}, Carlos},
        title = "{Dark matter halo merger histories beyond cold dark matter - I. Methods and application to warm dark matter}",
      journal = {Monthly Notices of the Royal Astronomical Society},
         year = 2013,
        month = jan,
       volume = {428},
       number = {2},
        pages = {1774-1789},
          doi = {10.1093/mnras/sts159},
archivePrefix = {arXiv},
       eprint = {1209.3018},
 primaryClass = {astro-ph.CO},
       adsurl = {https://ui.adsabs.harvard.edu/abs/2013MNRAS.428.1774B}
}

@ARTICLE{1993MNRAS.262..627L,
       author = {{Lacey}, Cedric and {Cole}, Shaun},
        title = "{Merger rates in hierarchical models of galaxy formation}",
      journal = {Monthly Notices of the Royal Astronomical Society},
         year = 1993,
        month = jun,
       volume = {262},
       number = {3},
        pages = {627-649},
          doi = {10.1093/mnras/262.3.627},
       adsurl = {https://ui.adsabs.harvard.edu/abs/1993MNRAS.262..627L}
}

@article{governato2007forming,
  title={Forming disc galaxies in $\Lambda$CDM simulations},
  author={Governato, Fabio and Willman, Beth and Mayer, L and Brooks, A and Stinson, G and Valenzuela, O and Wadsley, J and Quinn, T},
  journal={Monthly Notices of the Royal Astronomical Society},
  volume={374},
  number={4},
  pages={1479--1494},
  doi={10.1111/j.1365-2966.2006.11266.x},
  year={2007},
  publisher={Blackwell Publishing Ltd Oxford, UK}
}

@article{del2017small,
  title={Small scale problems of the $\Lambda$CDM model: a short review},
  author={Del Popolo, Antonino and Le Delliou, Morgan},
  journal={Galaxies},
  volume={5},
  number={1},
  pages={17},
  doi={10.3390/galaxies5010017},
  year={2017},
  publisher={Multidisciplinary Digital Publishing Institute}
}

@article{marinacci2014formation,
  title={The formation of disc galaxies in high-resolution moving-mesh cosmological simulations},
  author={Marinacci, Federico and Pakmor, R{\"u}diger and Springel, Volker},
  journal={Monthly Notices of the Royal Astronomical Society},
  volume={437},
  number={2},
  pages={1750--1775},
  doi={10.1093/mnras/stt2003},
  year={2014},
  publisher={Oxford University Press}
}

@article{paduroiu2022warm,
  title={Warm dark matter in simulations},
  author={Paduroiu, Sinziana},
  journal={Universe},
  volume={8},
  number={2},
  pages={76},
  doi={ 
10.3390/universe8020076},
  year={2022},
  publisher={Multidisciplinary Digital Publishing Institute}
}

\begin{appendix}

\section{Convergence Properties of Galacticus Merger Trees}
\label{sec:AppA}

\begin{figure*}[hbt!]
    \includegraphics[width = \linewidth]{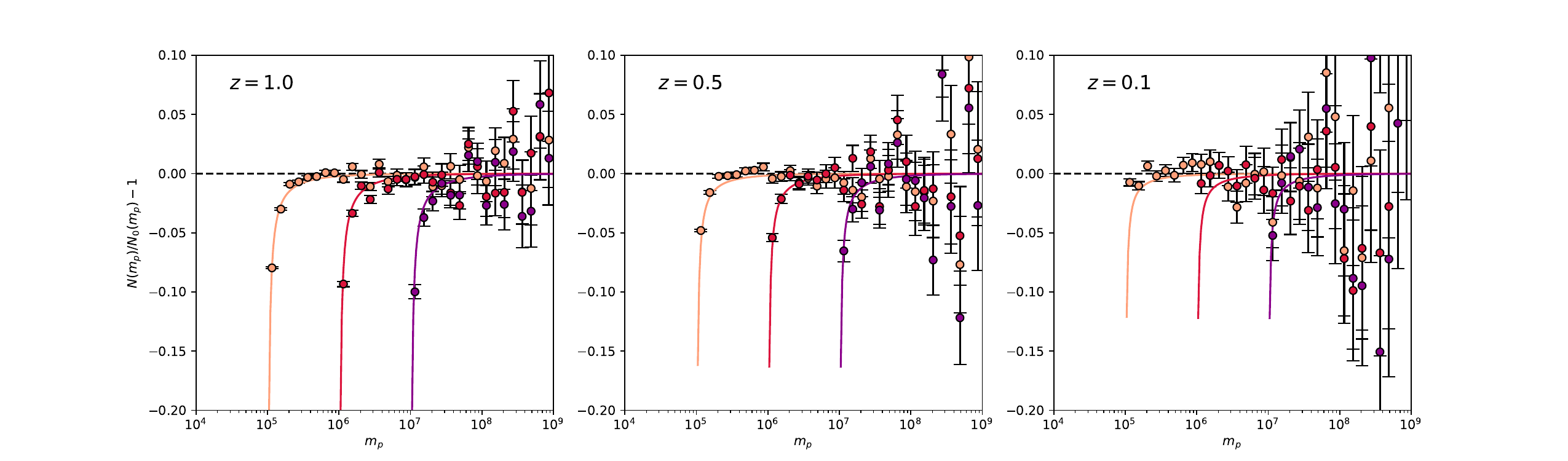}
    \centering
    \caption{The fractional offset of progenitor mass functions---for $ z = 1 $ (left), $ z = 0.5 $ (middle), and $ z = 0.1 $ (right) panels---relative to a very high-resolution ($10^4\mathrm{M}_\odot$) reference model, computed from {\sc Galacticus} merger trees. The dashed $ y = 0 $ line indicates the line of perfect convergence. Orange, red, and purple points indicate {\sc Galacticus} data at mass resolutions of $ 10^5$ (orange), $10^6$ (red), and $10^7 \mathrm{M}_\odot $ (purple), respectively, with $ 1\sigma $ uncertainties displayed. Solid lines show the fitting function given by equation~(\ref{eq:app}).}
    \label{PMF}
\end{figure*}

We evaluate the convergence behavior of {\sc Galacticus} merger trees using a sample of 240 realizations generated using the Monte Carlo algorithm of \cite{parkinson2008generating}, assuming a host halo mass of $10^{12} \mathrm{M}_\odot$. Figure \ref{PMF} shows fractional offsets in progenitor mass functions (PMFs) $ N(m_p) $ computed at mass resolutions of $ m_\text{res} = 10^5, 10^6, 10^7 \mathrm{M}_\odot $ relative to $ N_0(m_p) $, which denotes the PMF of a very high resolution model at $ m_\text{res} = 10^4 \mathrm{M}_\odot $. On the vertical axis in Figure \ref{PMF}, the dashed $ y = 0 $ line corresponds to exact convergence of a given model to the $ m_\text{res} = 10^4 \mathrm{M}_\odot $ reference model. Different panels correspond to different redshifts as indicated in the panel, and colored data points correspond to {\sc Galacticus} PMFs at varying mass resolutions. Specifically, orange, red, and purple data points correspond to $ m_\text{res} = 10^5, 10^6, 10^7 \mathrm{M}_\odot $ resolutions, respectively, with corresponding $ 1\sigma $ error bars.

It can be seen that, well above the resolution limit of each model, the results converge to those obtained from our highest resolution reference model. As the mass resolution of a given model is approached, the progenitor mass function contains fewer halos than the high-resolution reference model, indicating a lack of convergence. Quantitatively, the number of progenitor halos is underestimated by around 2\% at masses of $2 m_\mathrm{res}$, with the underestimate getting rapidly worse at lower masses. 

To model this convergence behavior, we fit a simple analytic function to these results:
\begin{equation} \label{eq:app}
    \frac{N(m_p)}{N_0(m_p)} = \exp \left( - \frac{4.93 \times 10^{-3}w}{[m_p/m_\text{res} - 1]^{0.871}} \right),
\vspace{1mm} 
\end{equation}
where $ w = \delta_c(z)/D(z) $ is the ratio between the collapse threshold and the linear growth factor evaluated at the relevant redshift. This fitting function is shown by the solid curves in Figure~\ref{PMF}.

In this work, we run {\sc Galacticus} at a mass resolution of $ m_\text{res} = 3 \times 10^7 \mathrm{M}_\odot $. Summary statistics of {\sc Galacticus} halo populations throughout the paper adopt a minimum halo mass of $ m = 1.2 \times 10^8 \mathrm{M}_\odot $ to match the halo mass resolution of $N$-body simulations. Therefore, the {\sc Galacticus} halos included in our analysis are at least 4 times more massive than the mass resolution. By examining the plots in Figure \ref{PMF}, we see that the progenitor mass functions of halos above 4 times the mass resolution are converged to better than $ 1\% $ of their actual values, and the relative convergence improves at lower redshifts.  Utilizing the fitting function above, we estimate convergence to better than 4\% even at $z=10$, and to better than 2\% at $z=3$ (the typical collapse epoch of the lowest mass subhalos we consider in CDM). As such, our merger trees are more than sufficiently well-converged for the analyses carried out in this work, given the statistical uncertainties present in the N-body datasets to which we compare.

\end{appendix}

\end{document}